\documentclass[twocolumn,showpacs,preprintnumbers,amsmath,amssymb,prd]{revtex4}

\usepackage{graphicx}
\usepackage{dcolumn}
\usepackage{bm}

\begin{document}


\title{Modeling Bose Einstein Correlations via Elementary Emitting Cells}

\author{Oleg Utyuzh}
 \email{utyuzh@fuw.edu.pl}
\author{Grzegorz Wilk}%
 \email{wilk@fuw.edu.pl}
\affiliation{%
The Andrzej So\l tan Institute for Nuclear Studies, Ho\.za 69,
00-681 Warsaw, Poland
}%
\author{Zbigniew W\l odarczyk}
 \email{wlod@pu.kielce.pl}
\affiliation{Institute of Physics, \'Swi\c{e}tokrzyska Academy,
                  \'Swi\c{e}tokrzyska 15, 25-405 Kielce, Poland
}%

\date{\today}

\begin{abstract}
We propose a method of numerical modeling Bose Einstein Correlations
by using the notion of Elementary Emitting Cells (EEC). They are
intermediary objects containing identical bosons and are supposed to
be produced independently during the hadronization process. Only
bosons in EEC, which represents a single quantum state here, are
subjected to the effects of Bose-Einstein (BE) statistics, which
forces them to follow a geometrical distribution. There are no such
effects between particles from different EECs. We illustrate our
proposition  by calculating a representative number of typical
distributions and discussing their sensitivity to EECs and their
characteristics.
\end{abstract}

\maketitle

\section{\label{sec:intro}Introduction}

In every high energy collision experiment, a vast number of
secondaries (mostly pions) is produced encoding valuable information
on the dynamics of the hadronization process. Because of their
complexity, such reactions can only be investigated by numerical
modeling using specialized codes, Monte Carlo event generators
(MCEG) \cite{GEN}. They all use positively defined probabilities to
describe the particle production process. This means that they
neglect off-diagonal terms in the corresponding density matrices
formally describing such processes. As a result they did not
properly implement all the correlations, in particular the quantum
mechanical correlations between identical particles (both of
Bose-Einstein type for identical bosons (BE) and of Fermi-Dirac type
for identical fermions (FD)). The whole collision process is
expressed only by the product of single particle distributions. This
makes the present MCEG {\it a priori} impossible to properly account
for results of many experiments (starting from \cite{GGLP}), which
explicitly show features usually attributed to correlations of this
type \cite{BEC}. To describe such experiments, one has to introduce,
in some way, effects of quantum statistics, opening thus a new
subject - {\it Bose-Einstein} (BEC) or {\it Fermi-Dirac}
correlations \cite{BEC}.

This subject already has a long and well documented history
\cite{BEC} but it is still far from complete. Its most specific
features one is faced with are the following:  In experiments of an
exclusive type (measuring all produced secondaries, as in
\cite{GGLP}) it is possible (at least in principle) to construct the
properly symmetrized wave function of {\it all} identical pions and
thus account for our inability to determine which particle is
emitted from which position in the source. In the simplest approach
using a plane waves representation of particles, a such wave
function for $n_{\pi}$-pion state, has the form \cite{zajc}:
\begin{eqnarray}
\Psi_{\left\{p\right\}}(\left\{x\right\};\left\{ r\right\})
=\frac{1}{\sqrt{n_{\pi}!}}\cdot
\sum_{\sigma}\exp\left[-i\sum_{j=1}^{n_{\pi}} p_j\,
r_{\sigma(j)}\right]
 \label{eq:nwave}
\end{eqnarray}
($\sigma(j)$ denotes the $j^{th}$ element of permutation of the
sequence $\{1,2,...,n_{\pi}\}$ and we sum over all $n_{\pi}!$
permutations in this sequence, including the identity permutation).
Points of detection, $x$'s, will vanish when calculating
probabilities, but points of production of secondaries, $r$'s, will
remain. Because experiments measure only momenta of particles, one
must integrate over $\{r_j\}$ using some multiparticle
spatio-temporal distribution function, $\rho(\{r_j\})$. This is
usually {\it assumed} to be factorizable, i.e., expressed by the
product of independent single particle distributions,
$\rho(\{r_j\})=\prod_j \rho(r_j)$ \cite{zajc}. Assuming further a
totally chaotic hadronizing source one gets the probability of the
$n_{\pi}$-pion state in the form of {\sl permanent} of the matrix
$||\Phi_{ij}||$,
\begin{eqnarray}
\mathcal{P}_{1,..,n_{\pi}}\!\!\!&=&\!\!\!\frac{1}{n_{\pi}!}\sum_{\sigma}\prod_{j}
\Phi_{j,\sigma(j)} = perm\, ||\Phi_{ij}|| = \nonumber \\ &\equiv&
\frac{1}{n_{\pi}!}
\begin{tabular}{||cccccc||}
& $\Phi_{1,1}$ & & $\cdots$ & & $\Phi_{1,n_{\pi}}$ \\
& $\vdots$ & & $\Phi_{j,j}$ & & $\vdots$ \\
& $\Phi_{n_{\pi},1}$ & & $\cdots$ & & $\Phi_{n_{\pi},n_{\pi}}$ \\
\end{tabular} \, , \label{eq:permanent}
\end{eqnarray}
with
\begin{equation}
\Phi_{ij}=\int e^{iq_{ij}r}  \rho(r) d^4 \{r\} ,\qquad q_{ij} = p_i-p_j,
\end{equation} \label{eq:Phi}
(which are all real if $\rho(r) = \rho(-r)$).

However, most of modern experiments are of an {\it inclusive} type
as one measures effect of BEC on limited samples of secondaries
only. The unobserved part of the system acts then as a kind of {\it
thermal bath} influencing measured samples of data \cite{FOOT1}.
Also the registered multiplicities are much higher prohibiting
calculation of permanents given by eq. (\ref{eq:nwave}). The
subsystem subjected to further experimental (and theoretical)
analysis forms only a (not too well defined) part of the total
system with strongly fluctuating characteristics to be averaged
during data collection procedure. Because, as we have mentioned,
such systems can be described only by numerical modeling, one faces
the fundamental question of how to account for BEC in such modeling
processes? It must be realized that this is a completely different
thing then a simple calculation of the usually considered
$n_{\pi}=2$ case, which in the case considered here amounts to a
well known expression for the $2-$particle BE correlation function
defined as a ratio of a two particle distribution and single
particle distributions with $Q=\vert p_i - p_j\vert $ ,
\begin{eqnarray}
C_2(Q) &=& \frac{N_2(p_i,p_j)}{N_1(p_i)\, N_1(p_j)}
             = 1 + \left | \int dr\rho(r)\cdot e^{iQr} \right |^2=  \nonumber\\
             &=& 1 + \mid \tilde{\rho}(q=Q\cdot R)\mid ^2 \label{eq:C2a}
\end{eqnarray}
($C_2 \rightarrow 2$ when $Q \rightarrow 0$ and $C_2 \rightarrow 1$
for large values of $Q$) \cite{BEC,rev,Z,Weiner}. In fact, in the
majority of cases, one is following precisely this route with
different functional forms of $\tilde{\rho}(QR)$ and sometimes even
with an entirely different form of $C_2$ (cf., for example,
\cite{Kozlov}), but always derived in some analytical way - both for
one and three dimensional cases. One is then fitting such
correlation functions to appropriate sets of data, aiming to get
information on the quantity $R$ (and similar others). The reason for
this is that $C_2(Q)$ is usually regarded as a (kind of) Fourier
transform of the space-time characteristic of the emitting source,
$\tilde{\rho}$, therefore providing information on $\rho(r)$
\cite{FOOT2}.

It is precisely this fact that makes BEC so interesting, but we
shall not follow this route here. Our aim is to provide an algorithm
that could cope with BEC from its first steps (to {\it model} it)
and which could bring this effect to other MCEGs once implemented
there. This we shall do by using the notion of Elementary Emitting
Cells (EEC) introduced some time ago \cite{BSWW}. They are some
intermediary objects (in fact quantum states) containing identical
bosons (henceforth we shall assume them being pions), which are
supposed to be produced independently during the hadronization
process. Only bosons in EEC are subjected to Bose-Einstein (BE)
statistics, which makes them follow a geometrical distribution.
There are no such effects between particles from different EECs.
Using this idea we propose an algorithm that can be used in any
MCEG, which has effects of BEC as one of its basic features. This
will be presented on the background of previous attempts to model
BEC discussed in \cite{zajc,MP,MPcells,Cramer,ITM}. We shall
illustrate its action by calculating a representative number of
typical distributions and discuss their sensitivity to EECs and
their characteristics. In this work no comparison with experimental
data is offered, we limit ourselves only to a thorough discussion of
our algorithm. It is supposed to be the main building block of any
serious MCEG aiming for a real comparison with data and including
therefore, for example, also the distribution of energies one is
supposed to hadronize or possible flows in the system, i.e.,
subjects which are out of the scope of this work. In particular it
will be most useful in cases where the hadronizing source is well
defined (as in $e^+e^-$ annihilation processes and other elementary
processes) or where the number of secondaries is exceptionally large
(in high multiplicity events).

For completeness, in the next Section (\ref{sec:basis}) we shall
provide a short overview of numerical modelling of BEC proposed so
far. This will put our investigation in proper perspective. Section
\ref{sec:algorithm} contains details of the proposed algorithm and
examples of results obtained from its simplest version. The last
Section contains conclusions and a discussion. Some specific
problems connecting with the method proposed are addressed in
Appendices \ref{sec:A} - \ref{sec:CC}.

\section{\label{sec:basis}History of numerical modeling of BEC}

\subsection{\label{sec:BEC_in_mod}Imitating BEC with existing MCEGs}

{\it Imitating} BEC means to use the original outputs of the
existing MCEG codes and change them in such way as to reproduce
measured $C_2(Q)$ (or other experimental characteristics of BEC).
Two approaches are used for this. In the first one one introduces
some special global weights (i.e., weights built for the whole
event) to bias accordingly the original results of MCEG
\cite{Weights} (usually checking whether other observables were not
changed too much - otherwise one has to re-run MCEG with new input
parameters). This procedure is justified by noticing the following
\cite{LUND}:  Let $M = \sum_{\sigma} M_{\sigma}$ be the matrix
element describing the production of a hadronic final state of $n$
identical bosons. It consists of $n!$ terms, each corresponding to a
particular permutation $\sigma$ of the $n$ identical particles in
the final state. In the simulation process of MCEG the interference
terms (off-diagonal elements in permanent eq. (\ref{eq:permanent}))
are neglected and one gets that the probability to produce such a
state is
\begin{equation} \left|M\right|^2_{MCEG} =
\sum_{\sigma}\left|M_{\sigma}\right|^2 \leq | M|^2.
\label{eq:weights1}
\end{equation}
To remedy this situation one {\it assumes} some weight
$\mathcal{W}_{\sigma}$ assigned to each event such that
\begin{equation}
\left|M'\right|_{MCEG}^2 = \sum_{\sigma}
\mathcal{W}_{\sigma}\left|M_{\sigma}\right|^2. \label{eq:weights2}
\end{equation}
There is no unique way to choose the weights $\mathcal{W}_{\sigma}$.
The only requirements is that one gets good fits to the
corresponding $C_2$ functions \cite{Weights,LUND}. The tacit
understanding is that then also $\left|M'\right|^2_{MCEG} \simeq |
M|^2$.

In the second approach, one locally modifies (by weighting each pair
of particles in a given event) the original output of the MCEG used.
This can be done either by modifying its energy momentum spectra
\cite{Shifts} or by changing the resulting charge assignment
\cite{UWW1,UWW2}. In the first case one introduces weight function
$f_{BE}(q)$ for pair of particles momenta which are changed, such
that
\begin{equation}
\int^Q_0 d \Omega(q) = \int^{Q+\delta Q}_0 f_{BE}(q)d\Omega(q)
,\label{eq:shift}
\end{equation}
i.e., for $f_{BE}(q) > 0$ one has $\delta Q <0$. In this way the
energy-momentum imbalance that results from such a procedure is
properly accounted for and number of particles, $N=\int d\Omega$, is
conserved \cite{Shifts}. In the second case \cite{UWW1,UWW2} the
original spatio-temporal and energy-momentum structure of the
original event is preserved, but often spurious unlike particle
correlations occur \cite{Unlike}. What must be stressed is that this
approach, contrary to the previous one, always works on the level of
a single event. It is therefore more suitable as an additional tool
(sometimes called {\it afterburner}) to be used together with the
known MCEGs.

It must be stressed that all these methods modify original physics
underlying MCEGs in an essentially unknown way \cite{FOOT4}. Using
them one {\it assumes} therefore that changes incurred are small and
irrelevant. We shall proceed now to attempts to {\it simulate} BEC
by which we understand a situation in which the algorithm
introducing effects of quantum statistics involves {\it all produced
particles} \cite{zajc,MP,Cramer,ITM}.

\subsection{\label{sec:Attempts_BEC}Attempts to simulate BEC numerically}

\subsubsection{\label{sec:zajc}Metropolis importance sampling method}

In \cite{zajc} the standard Monte Carlo technique due to Metropolis
({\it Metropolis importance sampling method}) was used. This is a
general method to generate an ensemble of $n$-body configurations
according to some prescribed probability density. In \cite{zajc}
this technique was used to modify the directions of momentum vectors
of selected particles from a system of $n$ identical particles in
order to impose the $n$-particle distributions derived from BE
correlation functions. In particular, it was done by changing the
momenta of selected particles, $p_i \rightarrow p'_i \in d^3N/dp^3$,
in such way as to maximize the probability of detection of the
$n_{\pi}$-multiparticle state, $\mathcal{P}_{1,...,p_{n_{\pi}}}$,
i.e. accepting a new configuration with probability
$\mathbb{P}=min\{1,P_{new}/P_{old}\}$, where
$P_{old}=\mathcal{P}\{1,...,p_i,...,p_{n_{\pi}}\}$ and
$P_{new}=\mathcal{P}\{1,...,p'_i,...,p_{n_{\pi}}\}$. This procedure
is then repeated many times until a kind of "equilibrium" is
achieved. As shown in \cite{zajc}, one was able in this way to
generate typical multipion events, which explicitly exhibit all
correlations induced by Bose statistics. The most important result
for our further consideration is the fact that as a result of the
application of this algorithm a number of objects, called {\sl
speckles} in \cite{zajc} and being clusters of a number of identical
pions in phase-space, is formed. It means that in the
multidimensional phase space permanent (\ref{eq:permanent}) exhibits
rich structure of local maxima (attracting particles) and voids
(repulsing them), which replaces the original distribution one
started from.  Actually the only drawback of this method is that
symmetrization of clusters with sizes larger than
$n_{cluster}\approx 10$ takes a prohibitively long time.

Two points must be stressed when summarizing this symmetrization
procedure. The first is that it involves all (identical) secondaries
in the event under consideration, some producing specific structure
in their distribution in the allowed phase space, namely it is
clustering them in some regions of phase space. The second point is
that this phenomenon leads immediately to a broadening of the
resultant multiplicity distribution (MD): starting from a Poisson MD
for a non-symmetrized wave function one ends up after symmetrization
with a geometrical (or Bose-Einstein) MD for a single speckle and
with Negative Binomial MD \cite{NBD} for the whole system.

We close this section with the following remark. So far particles
were represented, for simplicity, by plane waves. However, this
approach leads to some unpleasant effects because it violates the
Heisenberg uncertainty relation constraining the simultaneous
specification of coordinates and momentum as implied by
eq.(\ref{eq:C2a}). For example, in the case of sources with strong
position-momentum correlations the two-particle correlation function
$C_2(Q)$ can drop significantly below unity \cite{Weiner,Problems}.
This method has therefore been generalized in \cite{MP}, where plane
waves have been replaced by wave packets. Both features mentioned
above were also observed here. Therefore in \cite{MPcells}
modification simplifying the selection process was proposed. It was
argued that it could be limited only to identical particles whose
wave functions overlap in phase space, i.e., to particles forming
speckles or clusters mentioned above (with the size of this overlap
being a new parameter). Notice that such decomposition corresponds
to replacing the full permanent in eq. (\ref{eq:permanent}) by
matrix with a block structure, each block representing one cluster
with no correlations between them:
\begin{equation}
\begin{tabular}{||cccccc||}
& \fbox{$EEC_1$} & & $\cdots$ & & 0 \\
& $\vdots$ & & $\ddots$ & & $\vdots$ \\
& 0 & & $\cdots$ & & \fbox{$EEC_{N_{cell}}$}\, \\
\end{tabular} \,\,\, . \label{eq:cut-perm}
\end{equation}
In what follows we shall identify these blocks with EECs mentioned
before. However, it should be remembered that cells selected here
were so far preselected in $(\Phi,\Theta)$ space only and, by
construction, tend to contain particles with similar momenta. This
method differs therefore from that we are going to describe now in
which EECs are created dynamically without any restriction
\cite{Cramer}.

\subsubsection{\label{sec:cramer}The acceptance-rejection method}

The {\it acceptance-rejection method} used in \cite{Cramer} is the
well known "hit-or-miss" technique of generating a set of random
numbers according to a prescribed distribution, here given by an
expression describing the collapse of a multiparticle wave function
into a properly symmetrized state represented by
eq.(\ref{eq:permanent}), as required by Bose quantum statistics. In
contrast to that discussed above this method
\begin{figure}[!htb]
\vspace*{.3cm}
\begin{center}
\includegraphics*[width=6.0 cm]{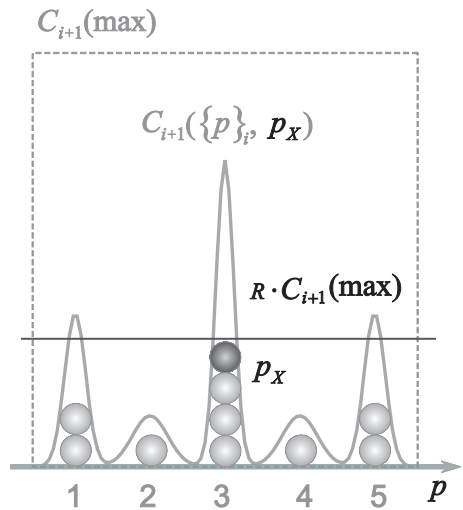}
\end{center}
\caption{Schematic example of acceptance-rejection algorithm
proposed in \cite{Cramer} at work; $C_{i+1}({\rm max})=(i+1)!$ and
$R\in [0,1]$ is random number. For $i$ particles already present
with $p_{i=1}$ fixed (gray circles, here $i=9$) one calculates
$C_{i+1}(\{p_{1,\dots,i}\},p_x=p_{i+1})$ and selects a random number
$R$. For the situation shown here (with schematic form of $C(p_x)$
as example) a new particle can only be added to "cells" $1$, $3$ and
$5$ but not to $2$ and $4$.} \label{Cramer}
\end{figure}
is sequential because the $n$ multiparticle event is constructed by
adding the $k^{th}$ particle to the $(k-1)$ particles already
selected, until $k=n$ is reached. This is done in the following way.
One starts with empty phase space and inserts in it the first
particle with momentum chosen according to some distribution (for
example, the one which is supposed to reproduce single particle
spectra). The presence of the first particle causes that a second
particle must be added according to the $2$-particle distribution
function, which for identical bosons is given by the $2$-particle
correlation function $C_2\propto P_{1,2}$ with momentum of the first
particle fixed and momentum of the second particle to be selected.
Because $C_2(Q=p_1-p_2)$ has a maximum at $Q=0$, the second particle
will tend to be located near the first one in phase space but {\it a
priori} it can take any position in it. Addition of a third particle
must be now performed according to $C_3 \propto P_{1,2,3}$ (with
momenta $p_1$ and $p_2$ already fixed and only $p_3$ being
selected), which again has bumpy structure, especially when
particles $1$ and $2$ are located far from each other. Therefore the
third particle will most probably locate itself near one of the
already selected particles but {\it a priori} it can take any
position in phase space. If, by chance, it will be far away from
both particles $1$ and $2$, it will become for future particles a
new point of attraction and seed of the new bump. Notice that if,
say $k$ particles occupy the same region of phase space, the
strength of the bump they form, which attracts other particles,
increases $k!$ times.  This process, schematically illustrated in
Fig. \ref{Cramer}, continues until all particles are used (their
number can be either preselected, in which case the initial energy
will vary, or can result from the procedure itself when initial
energy is fixed).

To summarize: this procedure leads again to some specific nonuniform
distribution of particles in the allowed phase space with some
cell-like structures (bumps) showing up. It results from the fact
that regions with some particles inside them already present will
have a bigger chance to attract a new particle. Unfortunately, this
sequential procedure is even more time consuming than the previous
one and therefore rather unpractical.

\subsubsection{\label{sec:japan}Information Theory approach}

The only workable example of MCEG with features of Bose-Einstein
statistics built in from the very beginning has been proposed in
\cite{ITM}. The known total energy $E_{tot}$ has there been
distributed among a given (mean) number of secondaries, $\bar{n} =
\bar{n}_{+} +\bar{n}_{-} + \bar{n}_{0}$ (where
$\bar{n}_{+}=\bar{n}_{-}$), with limited transverse momentum
parameterized by the mean value $\langle p_T \rangle $, and it was
done in such way as to reproduce data on both single particle
distributions and those for BEC as well. To this end Information
Theory (IT) method was used to obtain the most probable (and least
biased) formula for rapidity distributions, $dN/dy$, and
multiplicity distributions $P(n)$. It resembles the usual grand
canonical distribution but is more general than that because the
"temperature" $T$ and "chemical potential" $\mu$ occurring there are
now two lagrange multipliers obtained by solving the corresponding
energy and particle conservation constraints. To also get $C_2(Q)$
it turned out that it is enough to additionally divide the available
rapidity space into cells of fixed size $\delta y$ each and
assumption that each cell can contain only identical particles
(i.e., pions of the same charge in this case). It is remarkable
that, in addition to reproducing all single-particle characteristics
of the collision as well as backward-forward correlations, with this
step one gets at the same time also multiplicity distributions
$P(n)$ in Negative Binomial (NB) form \cite{FOOT5}, the proper
structure of the two particle BEC function $C_2(Q)$ and
intermittency signal \cite{FOOT6}. The distinctive feature of this
method is that it deals only with measured momenta of produced
secondaries, there is no trace on unmeasured spatio-temporal
structure present in other methods. If one now wants to somehow
deduce this information one has to treat $C_2(Q)$ in the same way
one is treating all experimental results on BEC (i.e., in fact one
has {\it to assume} that it is a kind of Fourier transform of the
hadronizing source $\rho(r)$ and perform its routine analysis as in
\cite{rev,FOOT2,WeinerC}).

The most important result of \cite{ITM} for us is demonstration that
the decisive factor leading to proper structure of the correlation
function $C_2(Q)$ was bunching property in the rapidity space
introduced there.

\subsection{\label{sec:zalewski}Quantum optical analogy}

Let us therefore concentrate for a moment on this bunching feature
of bosons introduced in \cite{ITM}. At first let us remind ourselves
that it has been noticed and discussed already many times
\cite{Zal,Pur,GN,MPcells} and was regarded as manifestation of
quantum statistics \cite{FOOT6a}. It is especially widely discussed
and used in quantum optics \cite{OPTICS} but it was also employed to
describe some aspects of the multiparticle production processes
\cite{BIYA,QS,BSWW}. This is because $C_2$ can be also regarded as
some measure of correlation of fluctuations. One uses here the fact
that
\begin{eqnarray}
\langle n_1 n_2\rangle &=&  \langle n_1\rangle \langle n_2\rangle
                     + \langle \left(n_1 - \langle n_1\rangle\right)
               \left(n_2 - \langle n_2\rangle\right)\rangle\nonumber\\
               &=&
                \langle n_1\rangle \langle n_2\rangle
                 + \rho \sigma(n_1)\sigma(n_2) \label{eq:COV}
\end{eqnarray}
(where $\sigma(n)$ is dispersion of the multiplicity distribution
$P(n)$ and $\rho$ is the correlation coefficient depending on the
type of particles produced: $\rho = +1,-1,0$ for bosons, fermions
and Boltzmann statistics, respectively). It means therefore that one
can write the two-particle correlation function (\ref{eq:C2a}) in
terms of the above covariances (\ref{eq:COV}) stressing therefore
its stochastic character \cite{UWW1,UWW2,F}:
\begin{eqnarray}
C_2(Q) &=& \frac{\langle n_i\left(p_i\right)\left[ n_j\left(p_j\right) -
\delta_{ij}\right]\rangle}  {\langle n_i\left(p_i\right)\rangle\langle
n_j\left(p_j\right)\rangle} = \nonumber\\
       &=& 1 + \rho \frac{\sigma\left(n_i\right)}
                             {\langle n_i\left(p_i\right)\rangle}
                        \frac{\sigma\left(n_j\right)}
                        {\langle n_j\left(p_j\right)\rangle} -
                        \frac{\delta_{ij}}{\langle n_i\left(p_i\right) \rangle }.
\label{eq:algor}
\end{eqnarray}
Notice that for geometrical (Bose-Einstein) multiplicity
distribution $\sigma^2\left(n_i\right) =\langle
n_i\left(p_i\right)\left[ n_i\left( p_i \right) - 1\right]\rangle$
(corresponding to bosons, $\rho =1$) one gets, for $i=j$, $C_2(Q=0)
=2$, i.e., its maximal allowed value. This bunching property has
already been used in our previous proposition of MCEG of the
"afterburner" type \cite{UWW1,UWW2} mentioned in Sec.
\ref{sec:BEC_in_mod}, it was realized in the form of EECs introduced
in \cite{BSWW}. Essentially the same idea will be the cornerstone of
our algorithm, which we shall now present. Identical pions will be
assumed to be subjected to BEC only when inside EEC, those from
different EECs are totally independent (this can be also expressed
using notation {\it chaotic} and {\it coherent}, see Appendix
\ref{sec:A}).

\section{\label{sec:algorithm}Modeling Bose Einstein Correlations via
Elementary Emitting Cells - our algorithm}

\subsection{\label{sec:AlI}Introduction}

The lesson learned from the approaches presented in Sec.
\ref{sec:basis} is that construction of the proper quantum
multiparticle bosonic state can be performed  $(i)$ either by
symmetrization of the corresponding wave function constructed for
all produced particles \cite{zajc,Cramer,MP,MPcells} or $(ii)$ by
following quantum statistical arguments formulated in
\cite{Zal,Pur,GN} and directly bunching produced identical
secondaries with (almost) the same energy in phase space to form
EECs \cite{BSWW,ITM,MPcells}.  So far, the fact that the
distribution obtained this way is of NBD type was only shown in
\cite{zajc}. Nevertheless the emergence of bunching (both in
\cite{zajc} and \cite{Cramer}) is convincing and makes it an
essential characteristic of the bosonic character of produced
secondaries one has to account for. This will be our main assumption
in what follows.

We have decided to {\it model} the effects of proper symmetrization
of a multiparticle state by assuring that identical particles are
produced in bunches with geometrical distribution of particles. To
get such a distribution one has to choose particles sequentially
with some prescribed probability $\mathcal{P}$ until first failure.
If this failure happens for the $(N+1)^{th}$ trial one gets
immediately that
\begin{equation}
P(N)=(1-\mathcal{P})\mathcal{P}^N \quad {\rm with}\quad \langle N\rangle
= \frac{\mathcal{P}}{1-\mathcal{P}} .\label{eq:GBE}
\end{equation}
Notice that for $\mathcal{P}$ defined as
\begin{equation}
\mathcal{P} = \mathcal{P}_0\cdot \exp \left( -\frac{E_i}{T}\right)
\label{eq:PPP}
\end{equation}
(and only then) one gets the usual form of the Bose-Einstein
distribution for the $i^{th}$ EEC:
\begin{equation}
\langle N(E_i)\rangle \equiv \frac{1}{\mathcal{P}_0^{-1} e^{E_i/T}-1}.
\label{eq:BEE}
\end{equation}
Because, according to eq. (\ref{eq:PPP}) $\mathcal{P}_0$ controls
the maximal number of particles which can be allocated in a given
EEC, one can formally introduce a kind of "chemical potential" (as
in \cite{ITM}) defined as $\mu = T\ln \mathcal{P}_0$ and get
\cite{FOOT8}
\begin{equation}
\langle N(E_i)\rangle  = \frac{1}{e^{\frac{E_i - \mu}{T}}-1}.
\label{eq:BEE1}
\end{equation}
Such a choice of $\mathcal{P}$ is therefore very crucial in the
process of further modeling BEC and is thus the cornerstone of our
algorithm.

\begin{figure}[!htb]
\begin{center}
\includegraphics[width=2.8cm]{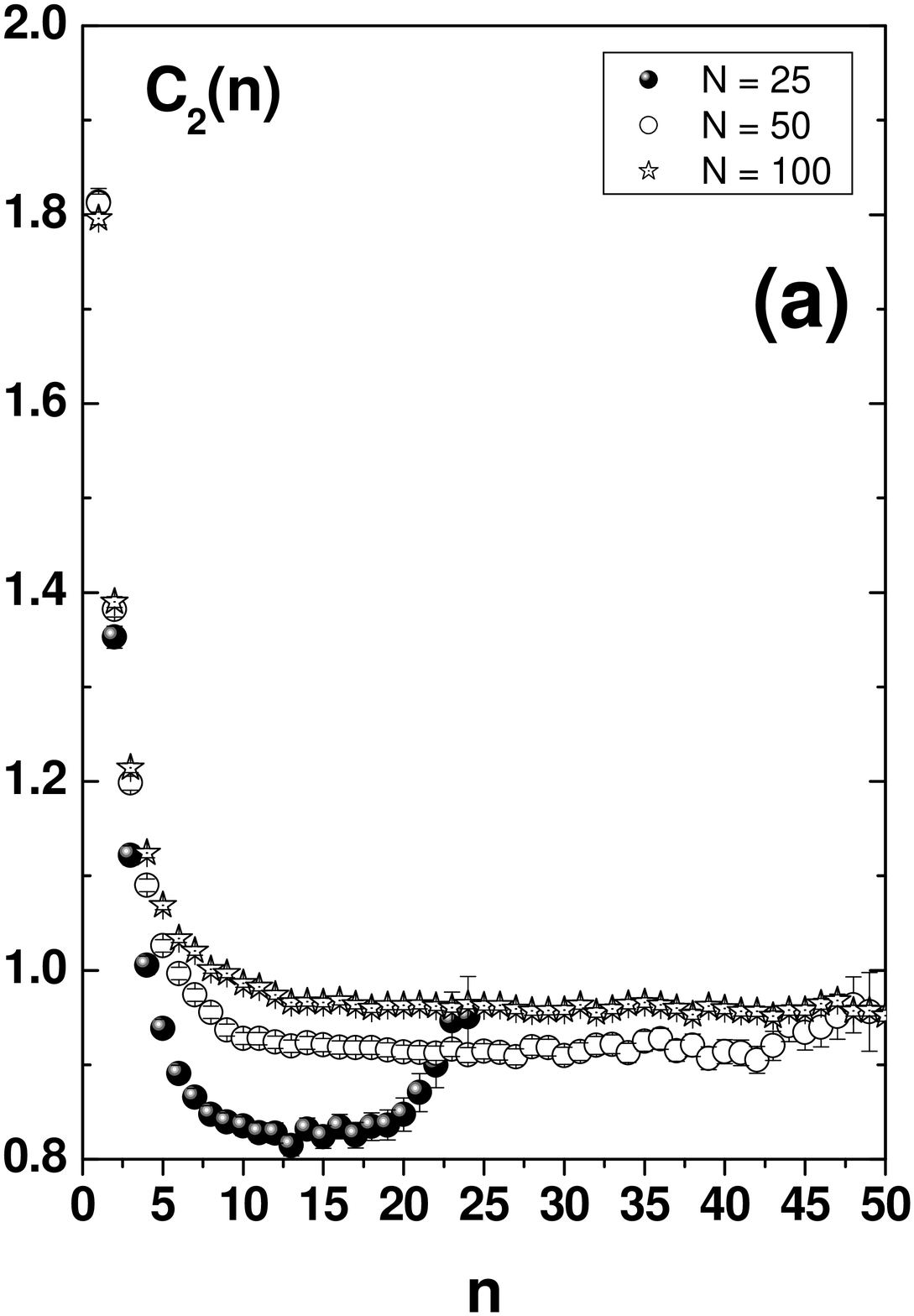}\hfill
\includegraphics[width=2.8cm]{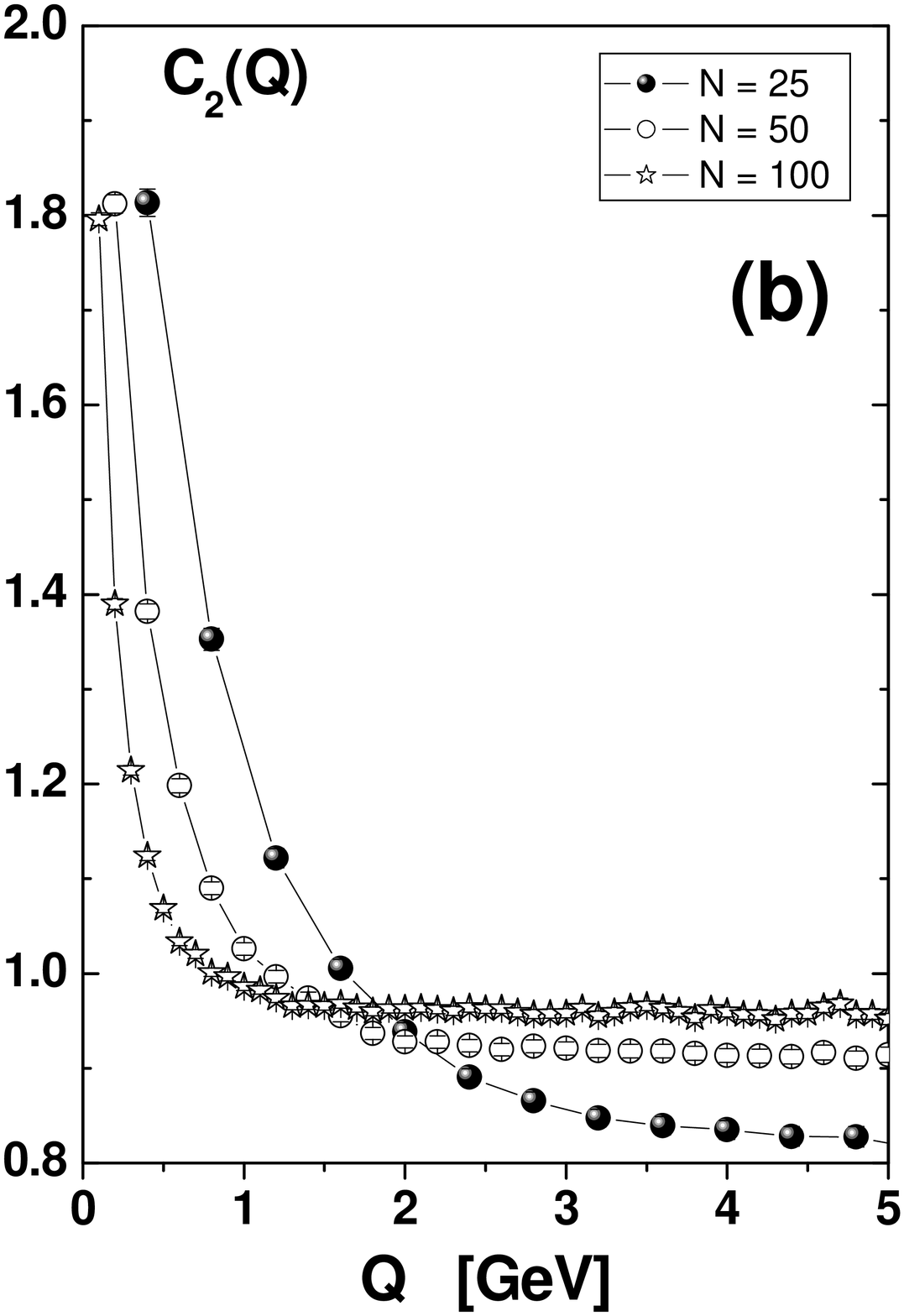}\hfill
\includegraphics[width=2.8cm]{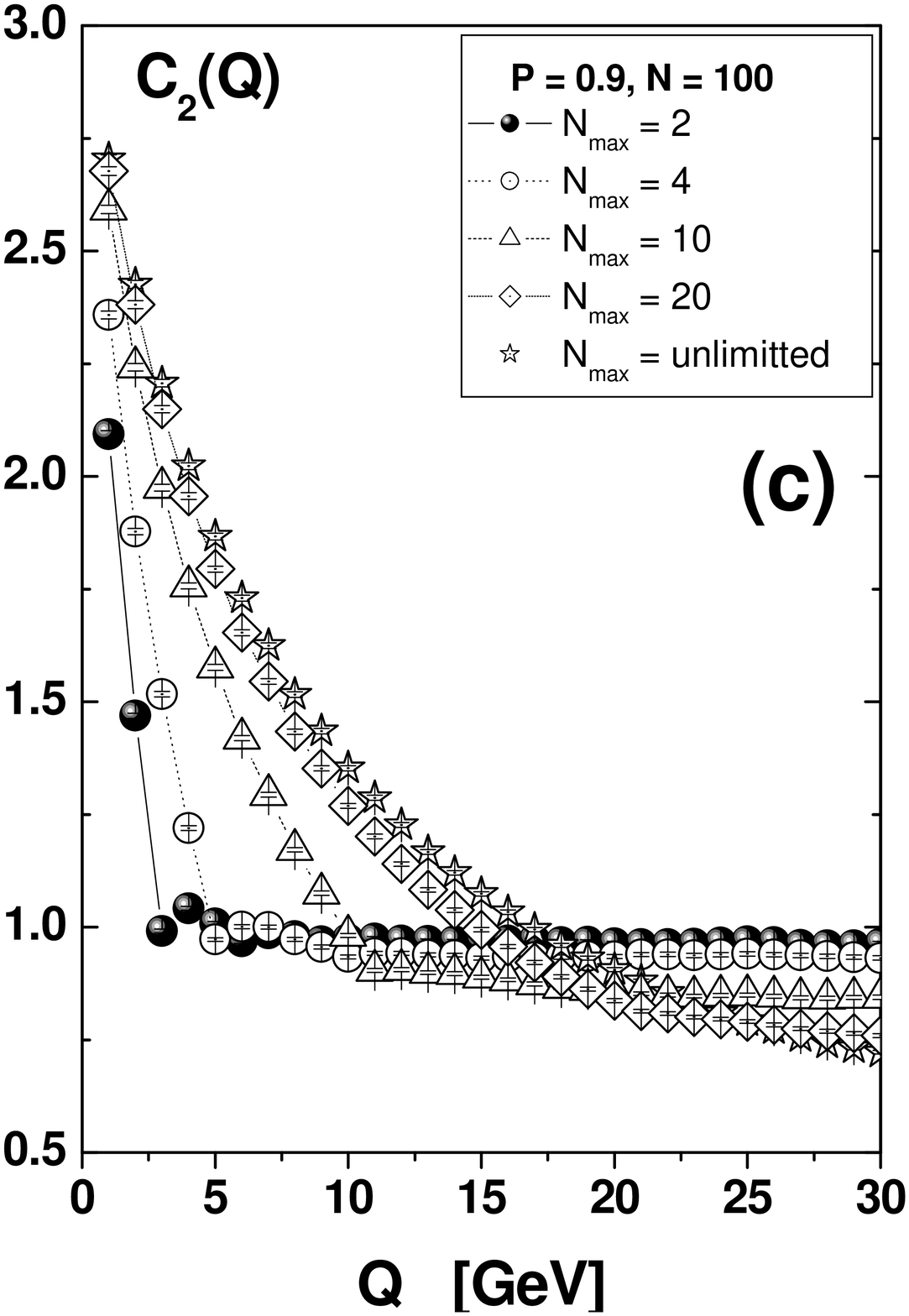}
\end{center}
\vspace{-0.6cm} \caption{Example of $C_2$ occurring for a pionic lattice
in (one-dimensional) momentum space:  $(a)$ $C_2(n)$ as function of
number of particles in a given cell for different $N$; $(b)$ $C_2(Q)$ as
function of $Q$ (in both cases  $\mathcal{P}=0.5$). In $(c)$ $C_2(Q)$ as
function of $Q$ defined by eq. (\ref{eq:exlat}) for artificially limited
cell occupancy, $i<n_{max}$ (now $\mathcal{P}=0.9$).} \label{fig:net}
\end{figure}

Let us now demonstrate this procedure on a simple example of a
one-dimensional pionic lattice model \cite{UWW2}. The pions located
in the sites of the lattice (which prevents identical pions to be
found at the same point of phase-space) are endowed with charges in
such way that after selecting the charge of the first pion, one
assigns the same charge to the consecutive neighboring pions with
probability $\mathcal{P}$ until first failure. After failure one
assigns (again randomly chosen) charge to the next pion in line and
continues this procedure until all pions are used. In this way a
number of cells is formed with particles in them distributed {\it by
construction} (cf. eq. {\ref{eq:GBE}) and \cite{FOOT8}) in
geometrical fashion, cf., Fig. \ref{fig:net}. In Fig. \ref{fig:net}a
dependence of $C_2$ on the lattice spacing defined by different
values of $N$ is presented whereas Fig. \ref{fig:net}b presents its
dependence on $Q$ defined as
\begin{equation}
Q=\mid p_i - p_j\mid= \frac{2p_{max}}{N}\cdot\mid i-j\mid=\delta p \cdot
n \label{eq:exlat}
\end{equation}
(where $p \in \left[-p_{max},p_{max}\right]$ with $p_{max} = 10$ GeV
and $N=100$ denotes the total number of sites in our lattice).
Comparing them one can deduce that the width of the correlation
function, $\sigma(C_2)$, is roughly proportional to the product of
the average number of particles in the cell, $\langle
N_{part}\rangle$, and the average distance $\langle\delta p\rangle$
between these particle on the lattice, $\sigma(C_2)\propto \langle
N_{part}\rangle\cdot \langle\delta p\rangle$. Because the average
distance between particles on the lattice is fixed by the (constant)
lattice spacing, the width of the correlation function depends only
on the mean cell occupancy, $\langle N_{part}\rangle$, which is
directly related to the probability $\mathcal{P}$ (cf.
eq.(\ref{eq:BEE})) and for constant $\mathcal{P}$ all widths of
$C_2(n)$ in Fig. \ref{fig:net}a are the same. Finally, in Fig.
\ref{fig:net}c the correlation function $C_2(Q)$ is shown for
different limits $N_{max}$ imposed on the maximal allowed cell
occupancy ($i<N_{max}$) (for $\mathcal{P}=0.9$ to assure large
occupancy of EECs). Notice that both the value of the intercept
parameter, $\lambda=C_2(Q=0)-1$, and the width of the correlation
function $C_2$ depend on the maximally allowed number of pions in
one cell, $N_{max}$.

\subsection{\label{sec:Impl}Implementation}

This experience prompted us to propose that BEC should be introduced
into the modeling procedure of multiparticle production processes as
early as possible, ideally at the very first stage of the selection
procedure. It means that one must devise some procedure how to
divide a given amount of energy $W$ into produced bosonic particles,
assumed to be pions of charges: $(+,-,0)$, without any {\it a
priori} assumption on what concerns their multiplicities other then
$n^{(+)}=n^{(-)}$ but with effects of quantum statistics accounted
for. Among procedures mentioned above only that in Sect.
\ref{sec:cramer} \cite{Cramer} satisfies this demand, however,
because it always involves all produced particles, it very quickly
becomes impossible to follow because of CPU time demand. On the
other hand,  results of \cite{zajc,Cramer} show explicitly that
quantum statistics leads to bunching particles in phase space. Such
bunching was therefore assumed in \cite{ITM,MP,MPcells} and in
\cite{ITM}, for the first time, the geometrical distribution form of
particles in bunches was used \cite{FOOT5}.

\begin{figure}[!htb]
\vspace*{.3cm}
\begin{center}
\includegraphics*[width=8 cm]{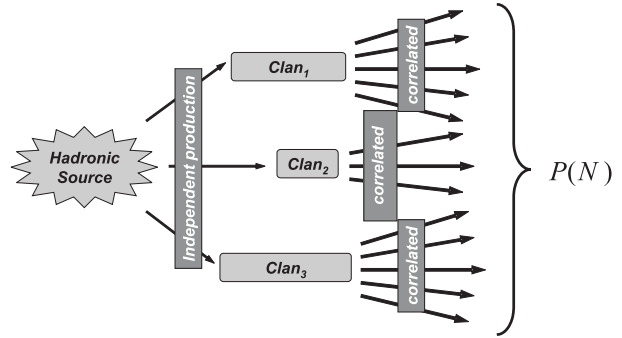}
\end{center}
\caption{Schematic view of {\it Quantum Clan Model} (QCM).} \label{QCM}
\end{figure}

We shall assume therefore that particles are produced according to a
mechanism which can be regarded as a quantum version of the {\it
clan model} (CM) introduced in \cite{NBD} to explain the fact that
apparently all multiparticle production processes result in the NB
form of resulting multiplicity distributions $P(N)$; we call it
therefore the {\it Quantum Clan Model} (QCM), cf. Fig. \ref{QCM}. In
CM model (distinguishable) particles are supposed to be produced in
clans, which were supposed to be formed independently (and therefore
distributed in Poisson fashion), with particles in clans following a
logarithmic distribution. Convolution of these two results in the NB
form of $P(N)$ ($\mathcal{P}$ denotes the probability of producing a
particle),
\begin{equation}
P_{NB}(N)=\left(
\begin{tabular}{c}
$N+k-1$ \\
$k-1$
\end{tabular}
\right)\mathcal{P}^k(1-\mathcal{P})^N . \label{eq:NBD}
\end{equation}
In QCM, because of quantum statistics, we must assume that each
quantum clan represents a single quantum state and therefore
contains only identical particles of (almost) the same energy, which
are distributed geometrically \cite{BSWW,Zal,ITM}. Keeping the
assumption of independent production of quantum clans, one
immediately finds that previous NB multiplicity distribution
(\ref{eq:NBD}) is now replaced by the so called P\'{o}lya-Aeppli
(Geometric-Poisson) distribution defined as \cite{PA} (with $\Theta
= (1-\mathcal{P})\langle N\rangle$):
\begin{eqnarray}
P_{PA}(N) &\equiv& Poisson\bigotimes Geometrical \label{eq:PA} \\
&=& e^{-\Theta}\mathcal{P}^{N}\sum_{j=1}^{N}\left(
\begin{tabular}{c}
$N-1$ \\
$j-1$
\end{tabular}
\right)\frac{\left[\Theta(1-\mathcal{P})/\mathcal{P}\right]^j}{j!},\nonumber
\end{eqnarray}
which was already used in multiparticle phenomenology some time ago
\cite{FD}. It differs from the NB only at small multiplicities.
Otherwise it is essentially the same.

\begin{figure}[!hbt]
\vspace*{0.3cm}
\begin{center}
\includegraphics*[width=4cm]{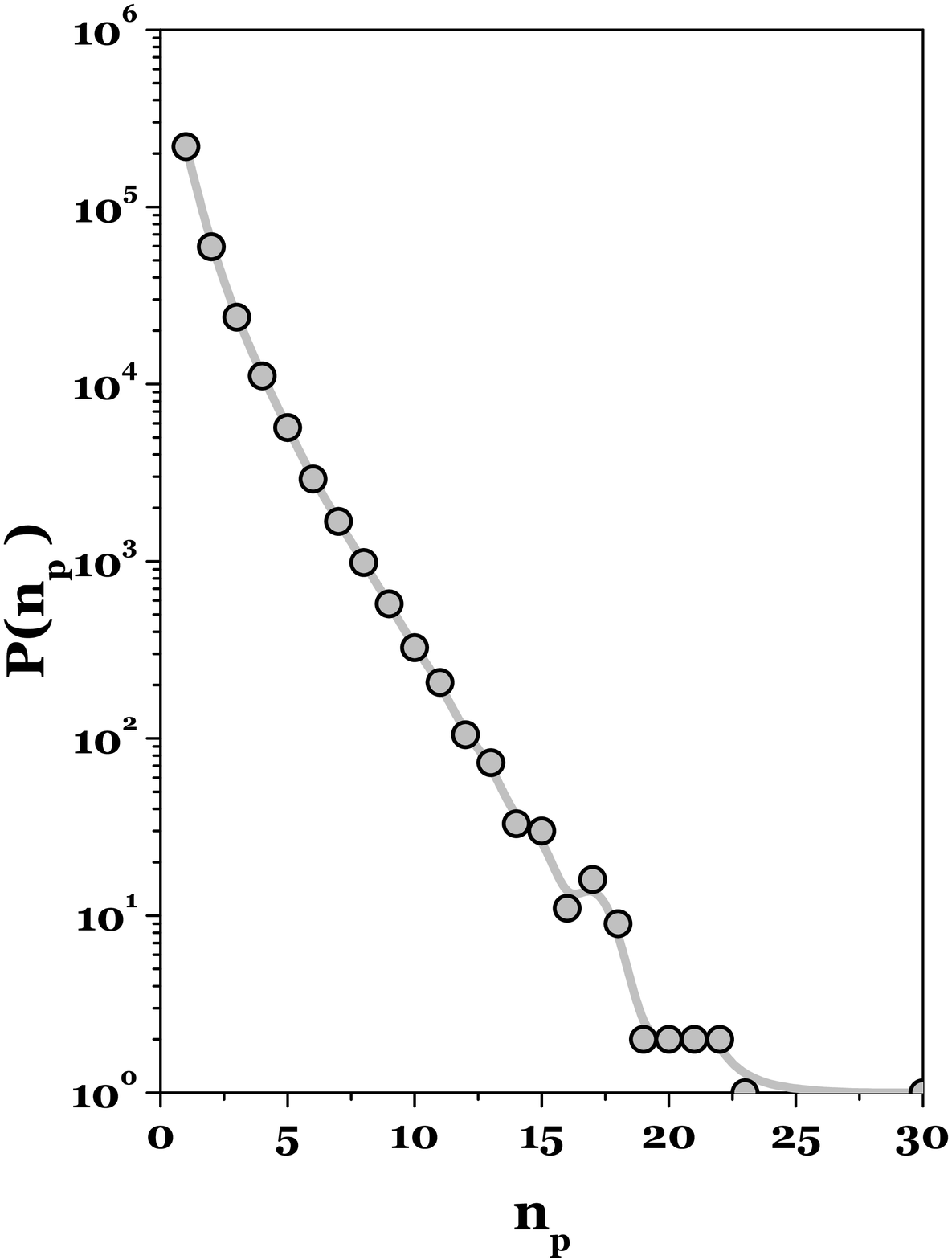}\hfill
\includegraphics*[width=4cm]{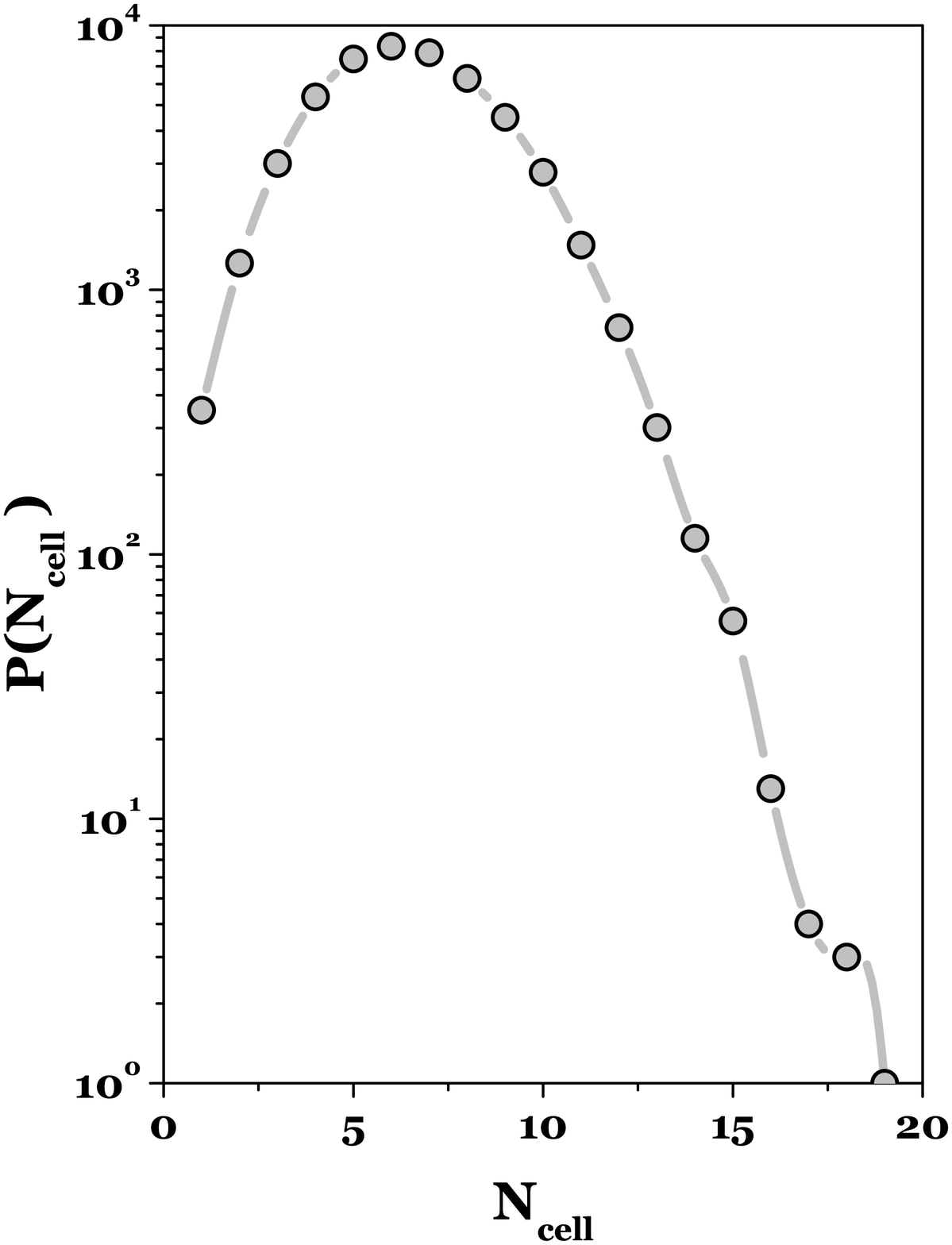}
\includegraphics*[width=6cm]{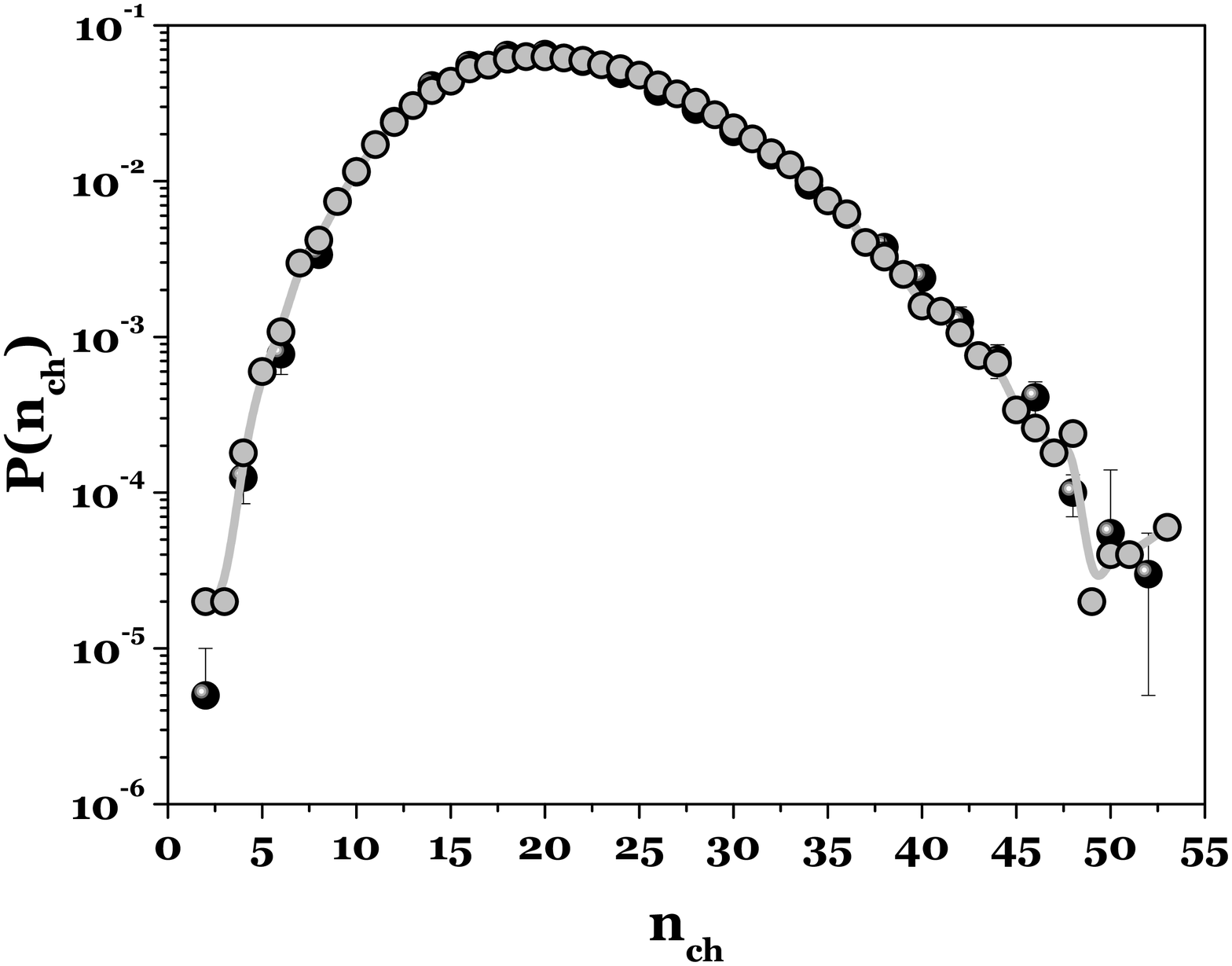}
\end{center}
\vspace*{-0.6cm} \caption{Examples of distribution of particles in a
given EEC (upper left panel), distribution of EECs (upper right
panel) and total charged particle distribution resulted from the
convolution of distributions shown in the first two panels (lower
panel, open circles and solid curve, see text for details). Notice
that, with this choice of parameters, one can reproduce exactly the
experimental data for $e^{-}e^{+}$ annihilations taken at the same
energy \cite{DELPHI} (black circles).}
\label{fig:cells}
\end{figure}

To implement numerically the QCM we proceed in the following way.

\noindent $(i)$ A pion of randomly chosen charge is selected with
energy $E_1 = E_{EEC}$ following some assumed distribution $f(E)$.
The form of this distribution is dictated by physics of the model
used to describe the production process which we would like to
follow (much the same as in \cite{Cramer}). It has no influence on
the bosonization (other then emerging from conservation laws), on
the other hand it strongly affects the final form of multiparticle
distributions obtained (by changing effective distributions of
EECs). This pion is therefore supposed to be the {\it seed} for the
first EEC.

\noindent $(ii)$ The other pions of the same charge are than added,
one after another, using probability $\mathcal{P}$ as given by eq.
(\ref{eq:PPP}). This is done until the first failure, which marks
the end of formation of the first EEC. After that, one starts
formation of another EEC by proceeding to point $(i)$ and choosing
from $f(E)$ another energy $E_{EEC}$ of another pion of randomly
chosen charge, which forms the seed of the second EEC. The whole
procedure is repeated until all available energy $M$ is used up.

\noindent $(iii)$ Once accepted, each of the selected pions forming
the first EEC is endowed with energy $E_i$ ($i\ge 2$), which is
selected from some distribution $G(E_i)$ centered on the energy of
the first pion forming the seed of this EEC, $E_{EEC}$.  The sense
of the function $G(E_i)$ is that it reflects the fact that the
requirement that all particles belonging to a given EEC must be in
the same energy state, in which case one expects that $G(E_i) \sim
\delta\left( E_{EEC} - E_i\right)$, means that their energies can
differ only by an amount corresponding to the width of the spectral
line mentioned in \cite{Pur}. We shall then use $G(E)$ either in the
form of delta function mentioned above or parameterize it by a
Gaussian,
\begin{equation}
G(E_i) = \frac{1}{\sqrt{2\pi}\sigma}\exp \left[ -
\frac{\left(E_{EEC}-E_i\right)^2}{2\sigma^2}\right]
.\label{eq:GaussE}
\end{equation}
In the thermal-like model for $f(E)$ used here, there is a natural
length scale given by the temperature $T$ and it is therefore
natural to choose $\sigma$ as being proportional to it, we shall
therefore take $\sigma=\sigma_0T$ (for the possible physical meaning
of $\sigma$ see Appendix \ref{sec:AA}).

\begin{table}[!hbt]
\caption{The mean total multiplicities and their dispersions ($\langle
N\rangle$ and $\sigma_N$), the mean total multiplicities of EECs  and
their dispersions ($\langle N_{cell}\rangle$ and $\sigma_{N_{cell}}$) and
mean multiplicities and dispersions in EEC ($\langle N_{part}\rangle $
and $~\sigma_{N_{part}}~$) calculated for  some selected choices of
parameters $\mathcal{P}_0$ and $T$ (upper part) and for different initial
energies ($W(i)=0.5W$, $W(ii)=W=91.2$ GeV and $W(iii)=2W$ - lower part).
All results are for $\sigma_0=0.0$ except the middle part, which was
obtained for $\sigma_0=0.3$ for comparison.}
\begin{center}
\begin{tabular}{||c|c||c|c|c|c|c|c||}
\hline
 & & & & & & &\\
 ~$\mathcal{P}_0$~ & $T$ & $~\langle N\rangle~$ & $~\sigma_N$ &
 $~\langle N_{cell}\rangle~$ & $~\sigma_{N_{cell}}$ &
$~\langle N_{part}\rangle~$ & $~\sigma_{N_{part}}$ \\
 & & & & & & &\\
\hline
 $0.9$ &       & 41.39 & 12.87 & 18.21 & 3.24 & 2.27 & 2.61 \\
 $0.7$ & $3.5$ & 33.81 & 8.28 & 20.58 & 3.73 & 1.64 & 1.28 \\
 $0.5$ &       & 30.34 & 6.63 & 22.41 & 4.11 & 1.35 & 0.78 \\
 $0.3$ &       & 28.22 & 5.69 & 24.01 & 4.44 & 1.17 & 0.48 \\
 \hline
       & $1.5$ & 82.68 & 15.29 & 39.85 & 4.66 & 2.07 & 2.13 \\
 $0.9$ & $3.5$ & 41.39 & 12.87 & 18.21 & 3.24 & 2.27 & 2.61 \\
       & $5.5$ & 27.99 & 11.13 & 12.03 & 2.64 & 2.32 & 2.78 \\
\hline\hline
 0.7 & 3.5 & 32.05 & 7.12 & 19.54 & 3.47 & 1.63 & 1.27 \\
\hline \hline
 & & & & & & &\\
 ~$W$~ & $T$ & $~\langle N\rangle~$ & $~\sigma_N$ &
 $~\langle N_{cell}\rangle~$ & $~\sigma_{N_{cell}}$ &
$~\langle N_{part}\rangle~$ & $~\sigma_{N_{part}}$ \\
 & $\mathcal{P}_0$ & & & & & &\\
\hline
 $(i)$ & $3.5$ & 21.22 & 9.08 & 9.49 & 2.29 & 2.24 & 2.56 \\
 $(ii)$ & $0.9$ & 41.39 & 12.87 & 18.21 & 3.24 & 2.27 & 2.61 \\
 $(iii)$ &       & 81.70 & 18.27 & 35.62 & 4.58 & 2.29 & 2.65 \\
\hline
\end{tabular}
\end{center}
\label{Parameters}
\end{table}

In this work we shall use the function $f(E)$ in the form of a
Boltzmann distribution,
\begin{equation}
f(E) = \exp\left( - \frac{E}{T}\right), \label{eq:PE}
\end{equation}
corresponding to a kind of thermal-like model with temperature $T$
being the main parameter. In this case the parameter
$\mathcal{P}_0$, governing the number of particles in EECs, plays
role of the chemical potential. As expected, one gets then EECs
distributed according to a Poisson distribution, each EEC containing
particles of the same charge only, which are distributed according
to Bose-Einstein (or geometrical) distribution, also the shape of
charged particle multiplicity distributions comes out as expected,
see Fig. \ref{fig:cells}. It was obtained for hadronizing energy $W
= 91.2$ GeV using  $T=3.5$ GeV, $\mathcal{P}_0=0.7$ and
$\sigma_0=0.3$. For the corresponding values of mean multiplicities
and their dispersion cf. Table \ref{Parameters} (the middle row).

The $W=91.2$ GeV value of energy will be used throughout the paper
(chosen to allow for the only comparison with data we show in Fig.
\ref{fig:cells}). In all examples shown here the number of MC trials
was $N_{MC} = 50000$ and reference frame used to calculate $C_2$
function is composed from $(+-)$ particles whereas $C_2$ themselves
are calculated for $(--)$ pairs. Table \ref{Parameters} shows the
corresponding multiplicities (and their dispersions) of all
particles obtained when hadronizing mass $W$ and those for the total
number of EECs and particles in them. The bigger $\mathcal{P}_0$,
the bigger the total multiplicity, smaller the number of EECs and
bigger their occupancy. Increase of $T$ decreases the total
multiplicity and number of EECs but slightly increases their
occupancy. The increase of available energy $W$ results in an
increase of all these quantities, except for the cell occupancy,
which remains essentially the same.

\begin{figure}[!hbt]
\begin{center}
\includegraphics*[width=4cm]{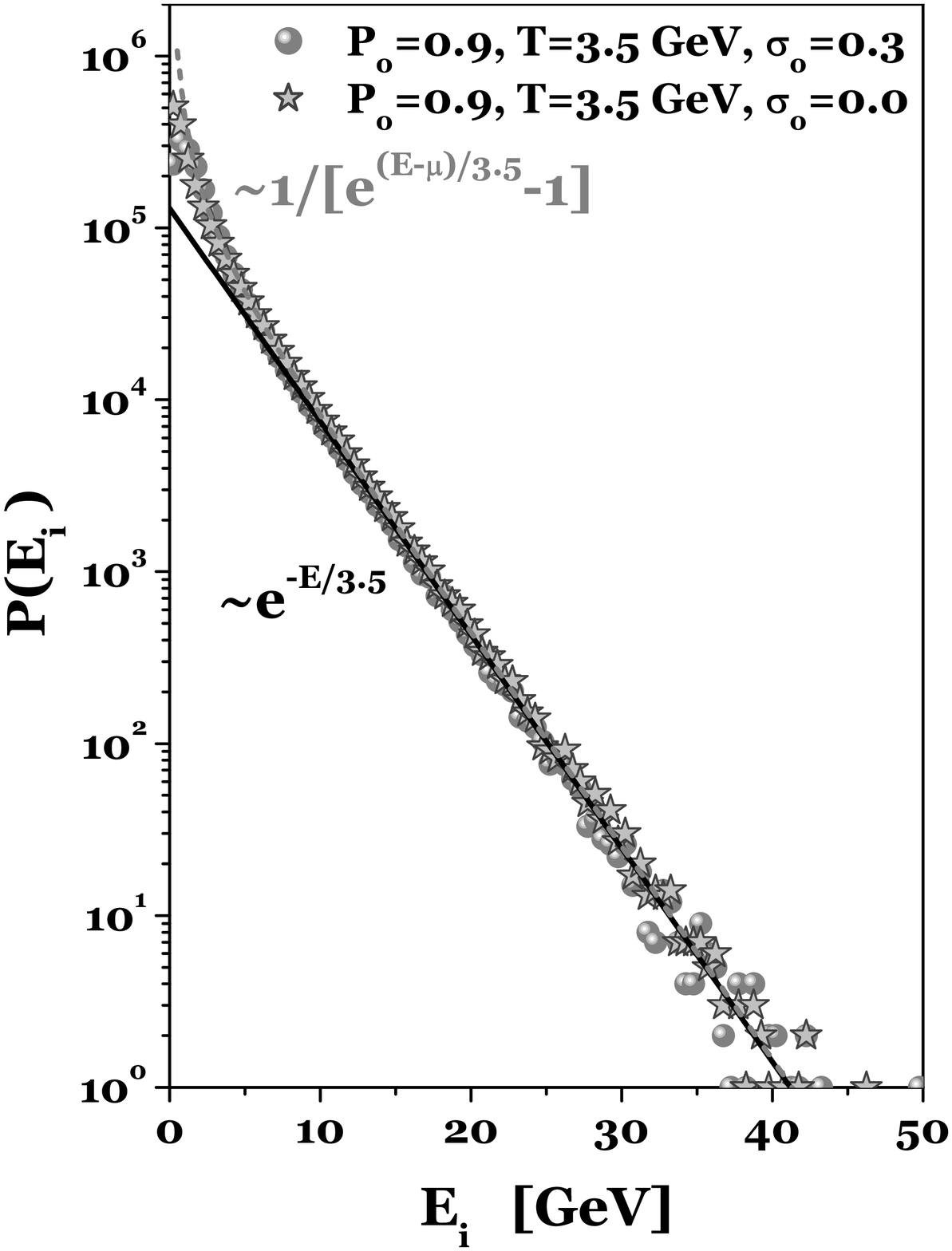}
\includegraphics*[width=4cm]{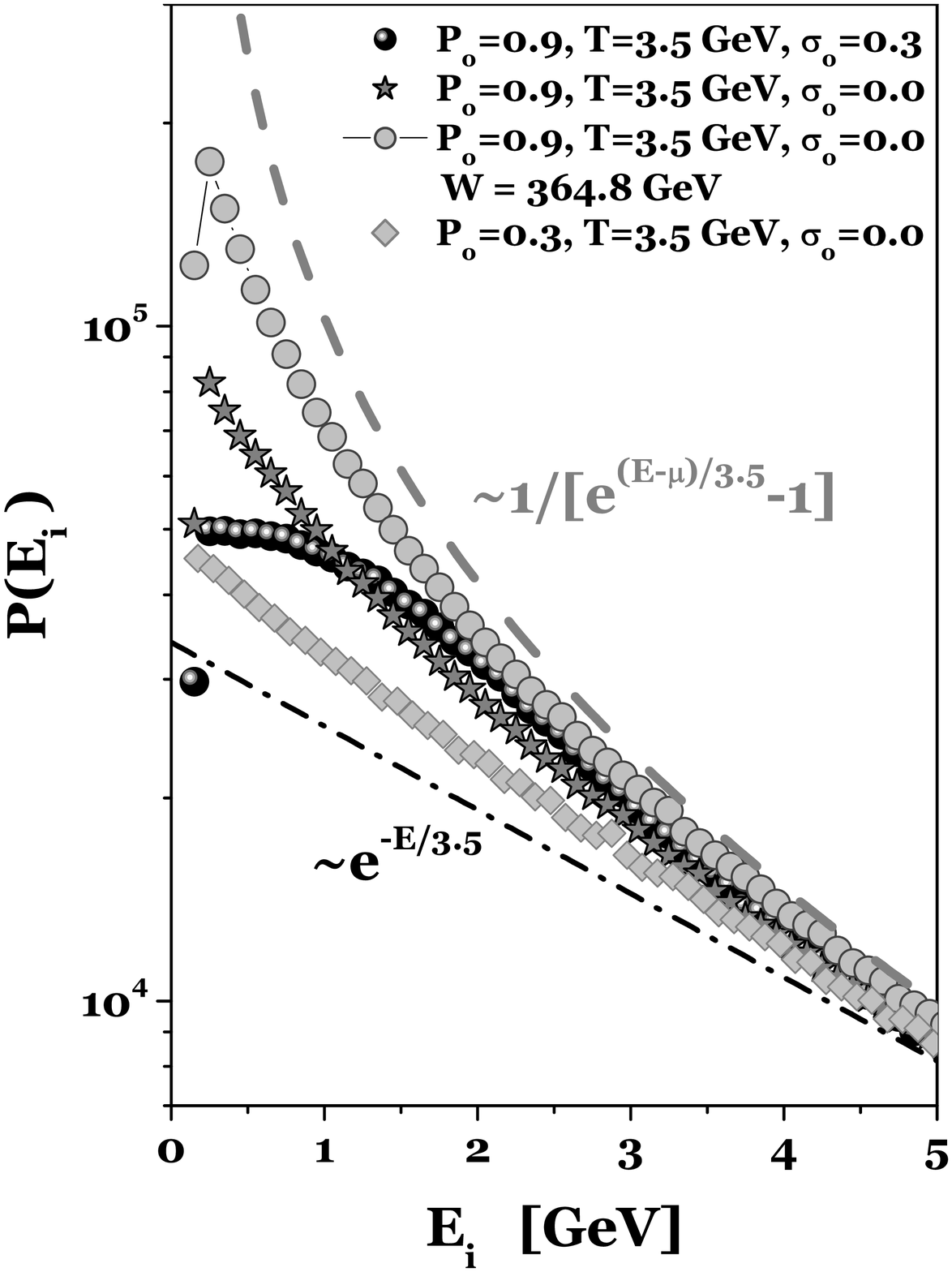}
\end{center}
\vspace{-0.5cm} \caption{Comparison of energy distributions obtained
using zero and nonzero values of $\sigma_0$ compared (dashed-line) with
the corresponding  Bose-Einstein form of energy dependence of occupation
number as given by eq. (\ref{eq:BEE}). The mass of hadronizing source is
$W = 91.2$ GeV, except the last curve where it is $4$-fold greater for
comparison.} \label{fig:Edist_s0} 
\end{figure}

Fig. \ref{fig:Edist_s0} shows an example of the corresponding energy
distributions of produced secondaries. Notice that the effect of
bunching (i.e., effect of introducing EECs) is visible only in the
limited range of the allowed phase space, concentrated at small
energies. In the case considered here, where the allowed range is
$(1/2)\cdot (W = 91.2)$ GeV, it practically vanishes for $E > 7.0$
GeV and after that value the distribution follows the exponential
form of $f(E)$ we have started from. This means that BEC increases
the multiplicity of event by adding particles with small energies
(see also Table \ref{Ecells}). Notice that for nonzero $\sigma_0$
one gets a displaced maximum for small values of $E$. At large
energies results follow the shape of the original $f(E)$
distribution (\ref{eq:PE}) used here. The other important feature is
the fact that none of the numerical simulations reproduces the
Bose-Einstein form of energy dependence of occupation number,
usually used in all analytical estimations and given by eq.
(\ref{eq:BEE}). This is because of the finiteness of available
energy $W$ one can use for hadronization, which results in limited
occupancy of EECs and violates conditions used to obtain eq.
(\ref{eq:BEE}) \cite{FOOT8}. Detailed results on the mean number of
EECs and charged particles multiplicity in them in different energy
bins are presented in Table \ref{Ecells}.

\begin{table}[h]
\caption{The mean number of EECs, $N_{cell}$, and charged particles
multiplicity in them, $n_{\pi^-}$, in different energy bins calculated
for $T=3.5$ GeV and two sets of $(\sigma_0, \mathcal{P}_0)$.}
\begin{center}
\begin{tabular}{||c||c|c||c|c||}
\hline
           &\multicolumn{2}{c||}{} & \multicolumn{2}{c||}{} \\
 Bins in $E$            &\multicolumn{2}{c||}{$\sigma_0 =0.1$; $\mathcal{P}_0=0.9$} &
            \multicolumn{2}{c||}{$\sigma_0 =0.3$; $\mathcal{P}_0=0.7$} \\
(GeV) &\multicolumn{2}{c||}{} & \multicolumn{2}{c||}{} \\
           &~~$N_{cell}~~ $   & $n_{\pi^-}$ &~~$N_{cell}$~~  & $n_{\pi^-}$       \\

\hline

 $0.0-0.5$    & 0.58& 2.62 & 0.63 & 1.2 \\

 $0.5-1.0$   & 0.72  & 3.00 & 0.79 & 1.60 \\

 $1.0-2.0$   & 1.16 & 2.98 & 1.26  & 2.62 \\

 $2.0-3.0$   & 0.87 & 1.58 & 0.95 & 1.70 \\

 $3.0-5.0$   & 1.15  & 1.64 & 1.25 & 1.76 \\

 $5.0-7.0$  &  0.65 & 0.77  &  0.71 & 0.83  \\

 $> 7.0$   &  0.84  &  0.89  &  0.92  &  0.97 \\

\hline
\end{tabular}
\end{center}
\label{Ecells}
\end{table}

\begin{figure}[!hbt]
\begin{center}
\includegraphics*[width=4.cm]{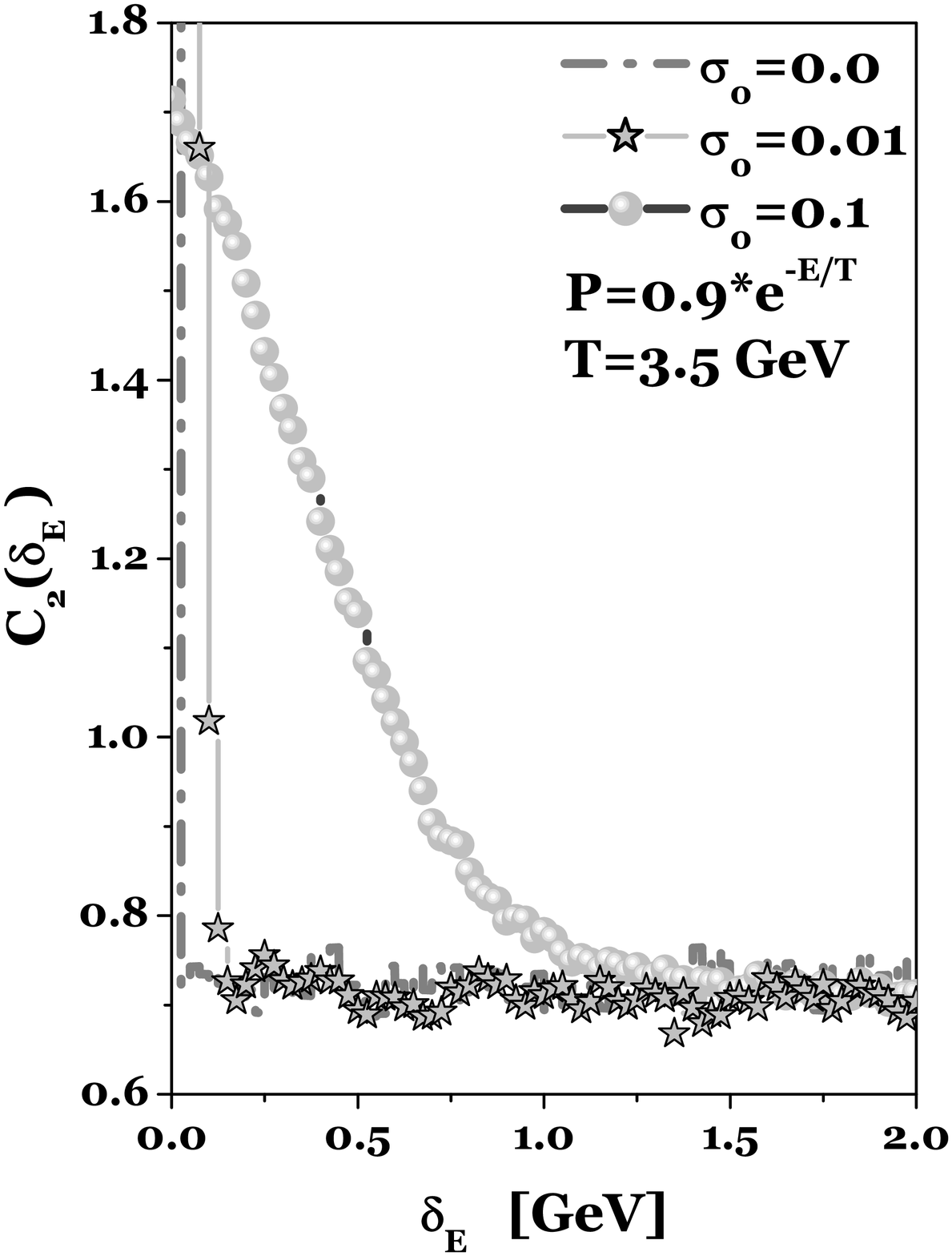}
\includegraphics[width=4.cm]{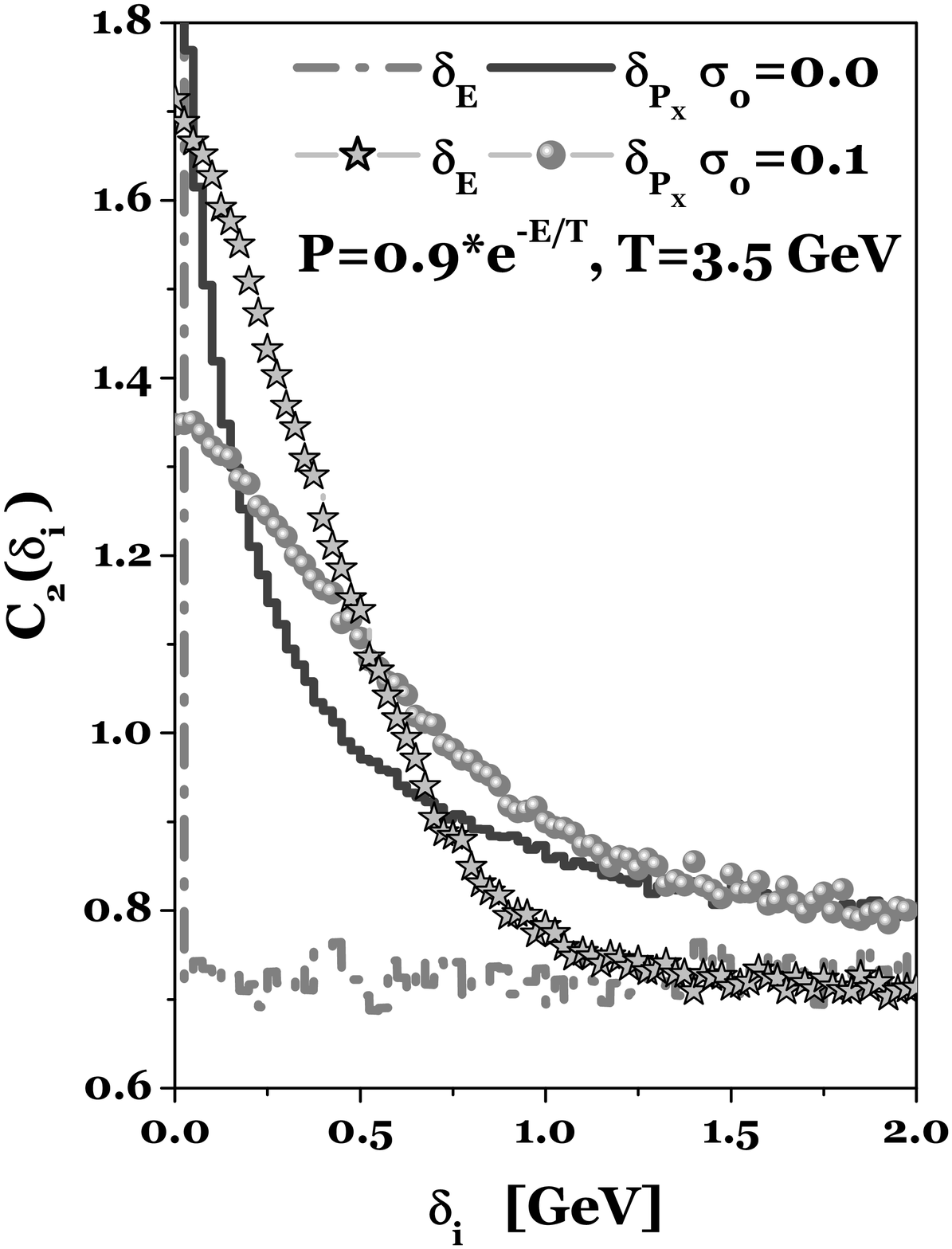}
\end{center}
\vspace{-0.6cm} \caption{Illustration of importance of spreading in
energy. Left panel: $C_2(\delta_E)$ case with of $\sigma_0=0.0$, $0.01$
and $0.1$. For $\sigma_0 = 0$ the maximum of $C_2$ is divergent (all
points are in the first bin). Right panel: $C_2(\delta_{E,p_x})$ cases
with $\sigma_0 = 0.0$ or $\sigma_0=0.1$. Notice that $C_2(\delta_{p_x})$
(calculated for momenta distributed isotropically) has nonzero width even
for $\sigma_0=0$. Introducing nonzero $\sigma_0$ results in further
broadening of $C_2$.} \label{fig:Kozlov} \vspace{-0.3cm}
\end{figure}

We now proceed to the correlation functions $C_2(\delta_X = X_1 -
X_2)$ and start with distributions in energy, $C_2(\delta_E)$ . In
Fig. \ref{fig:Kozlov} one observes that when using $\sigma_0=0$
(i.e., for strictly $\delta$-like form of $G(E)$) the whole effect
is located in the first bin only (this is just computer realization
of delta function). Therefore, if nonzero widths of $C_2$ are
needed, one must use $\sigma_0 >0$. Notice, however, that this width
is not equal to the input $\sigma = \sigma_0T$ used because the
difference of two variables, each following the same Gaussian
distribution, is again Gaussian but with twice $\sigma^2$, therefore
the final distribution should be $\sim \sqrt{2}$ broader, which is
roughly the case shown in Fig. \ref{fig:Kozlov} (where $\sigma =
0.35$ GeV was used as input in (\ref{eq:GaussE}) whereas the width
which can be read off from the obtained shape of $C_2$ is equal to
$0.45$ GeV).

We must stop for a moment to comment the fact that, in all figures
presenting $C_2(\delta_X)$ shown here, their values are
significantly smaller than unity for large values of $\delta_X$.
This is not an artifact of our algorithm, but results from the
method of presentation of our output. We want to keep the same
number of pairs both in the real event and in the reference one
(which, in our work, is always built from pairs of opposite
charges). Technically it means that one has to conserve the area
under each curve for $C_2(\delta_X)$. Actually this effect is also
known in all other approaches to BEC and is usually corrected by
arbitrarily shifting $C_2$ in such way that it equals unity for some
large value of the argument (in practice set to be equal $1$ GeV)
\cite{Weights}. We shall not do this here, but this fact must be
remembered when looking at our results.

In Fig. \ref{fig:Kozlov} $C_2(\delta_E)$ is compared with
$C_2(\delta_{p_x})$ calculated assuming an isotropic distribution of
momenta of particles in a given EEC, $\vec{p}$. The freedom
presented in the choice of directions of momenta results in a
nonzero width of the otherwise $\delta$-like structure of
$C_2(\delta_E)$ for $\sigma_0=0$. It then further broadens when one
allows for some nonzero width $\sigma_0$. This means that it can be
used as an additional parameter when comparison with data would be
attempted (provided its physical meaning will be made clear, see,
for example, the discussion in Appendix \ref{sec:AA} and
\cite{fractality}).

\begin{figure}[!htb]
\begin{center}
\includegraphics*[width=4.cm]{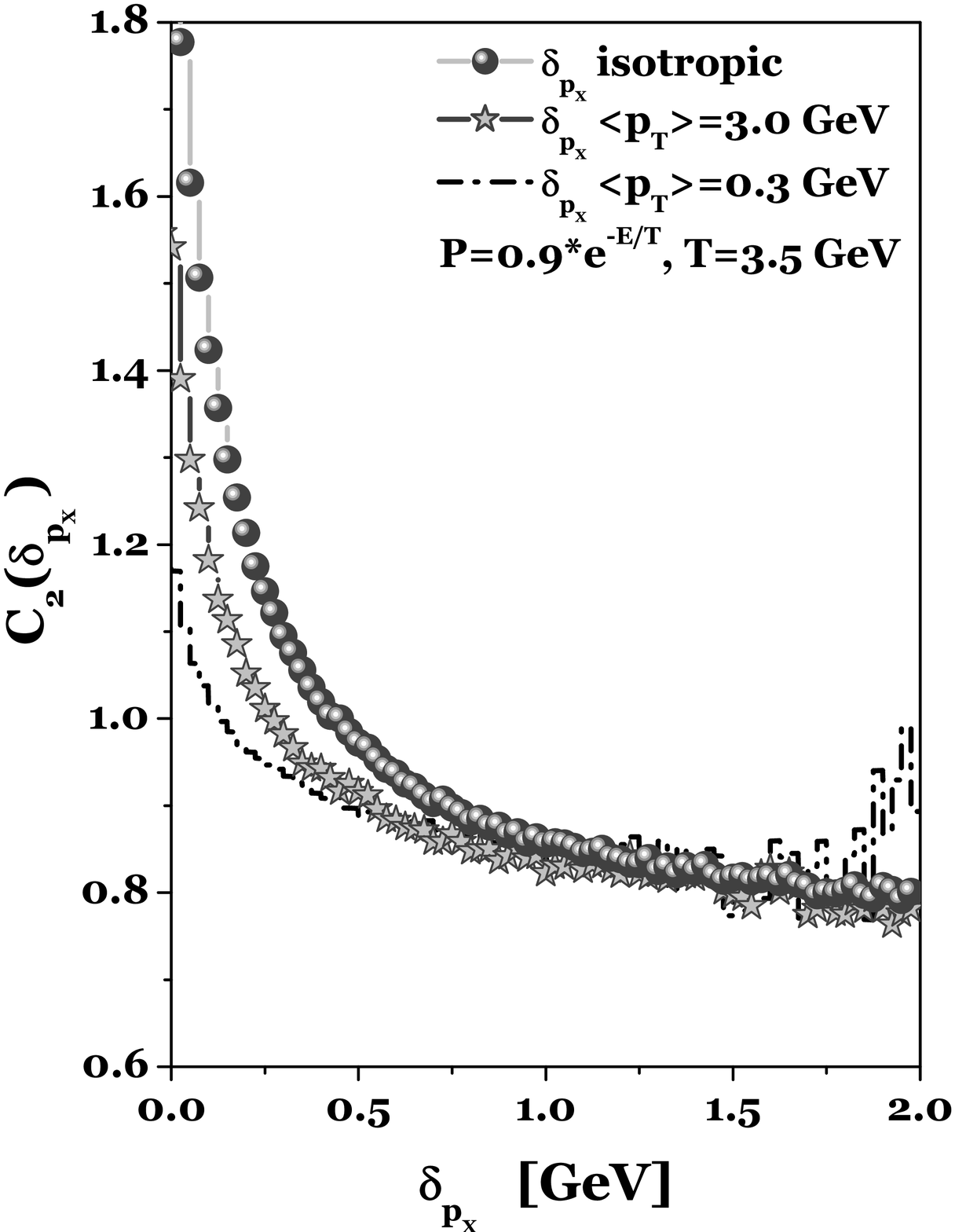}
\includegraphics*[width=4.cm]{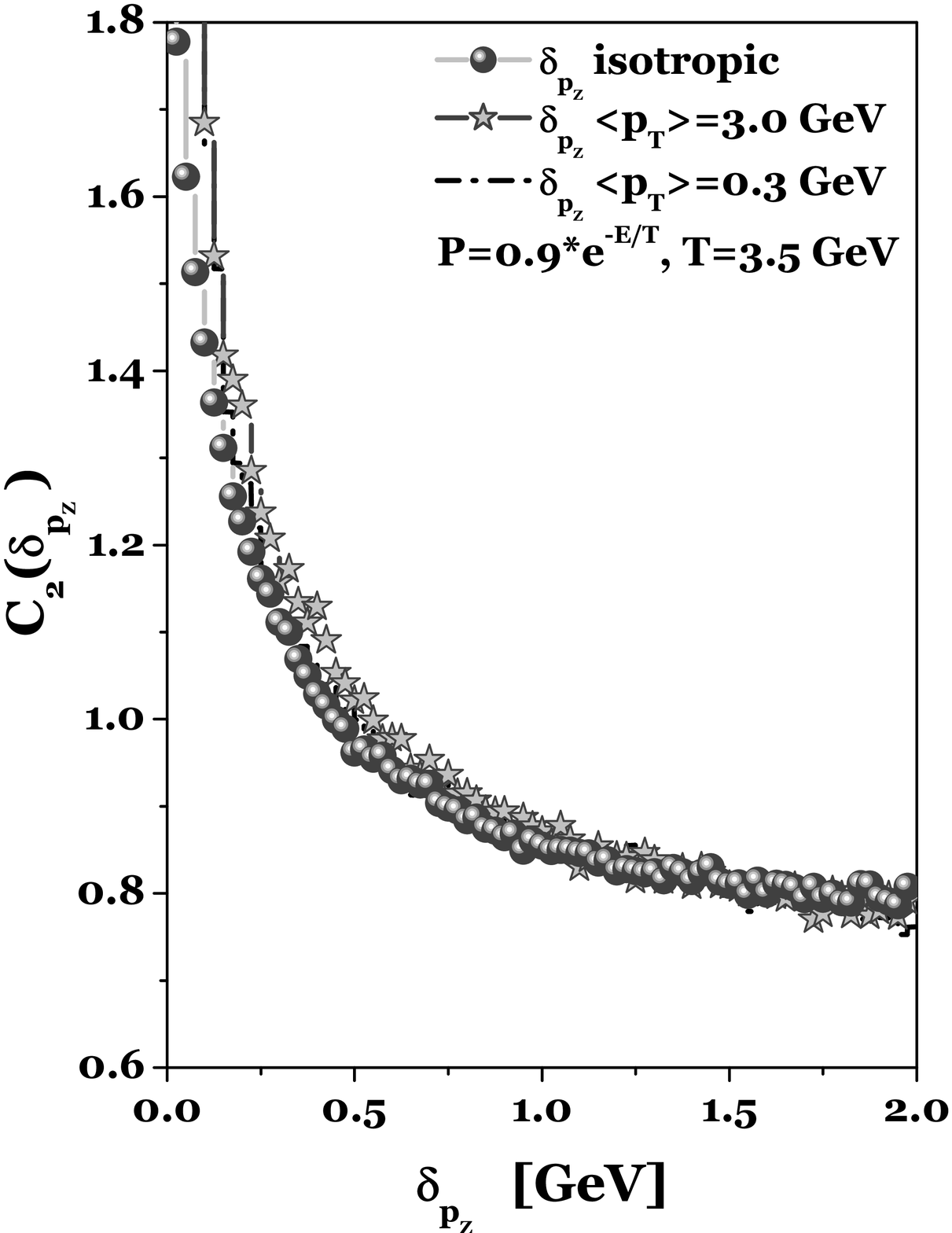}\\
\includegraphics*[width=4.cm]{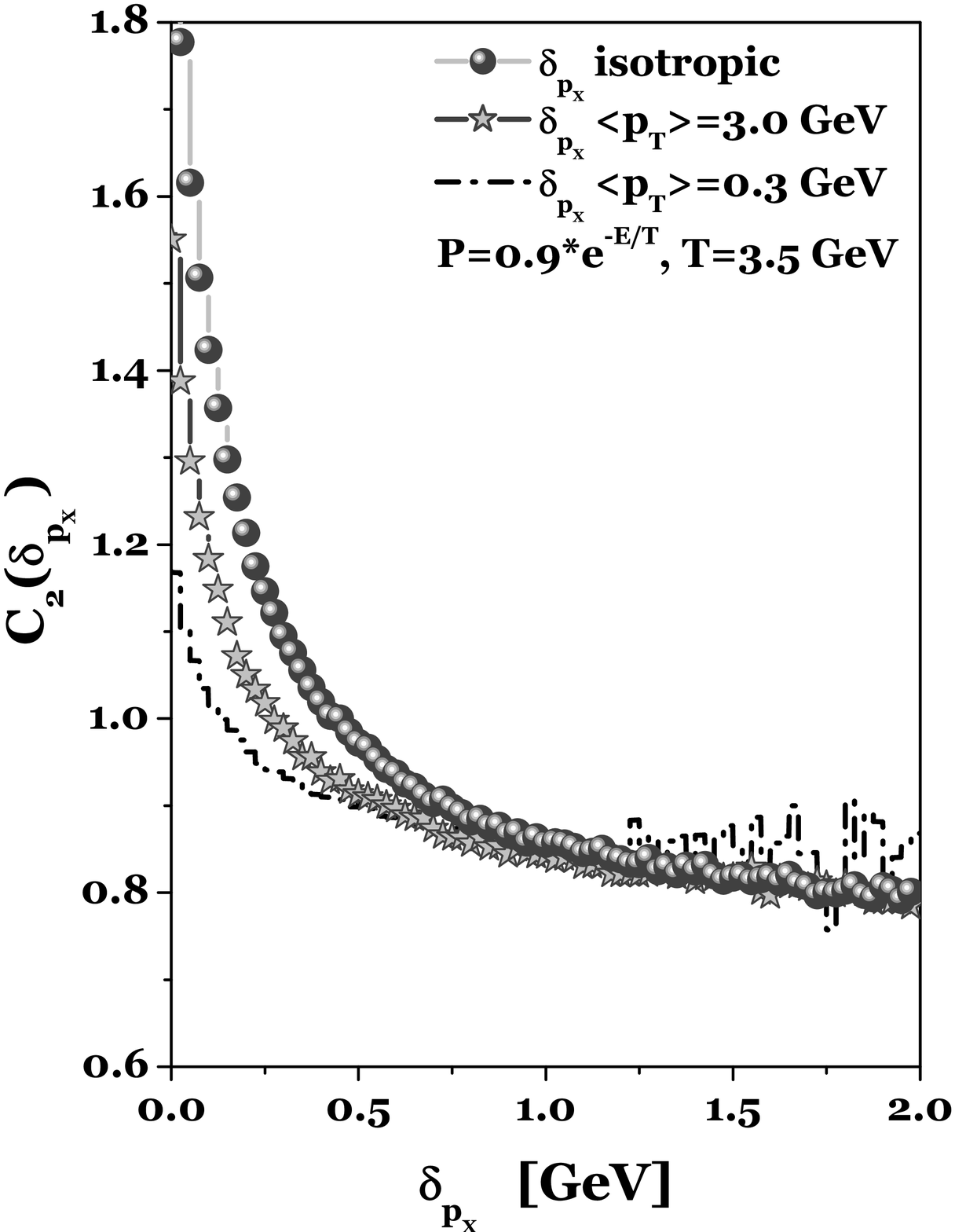}
\includegraphics*[width=4.cm]{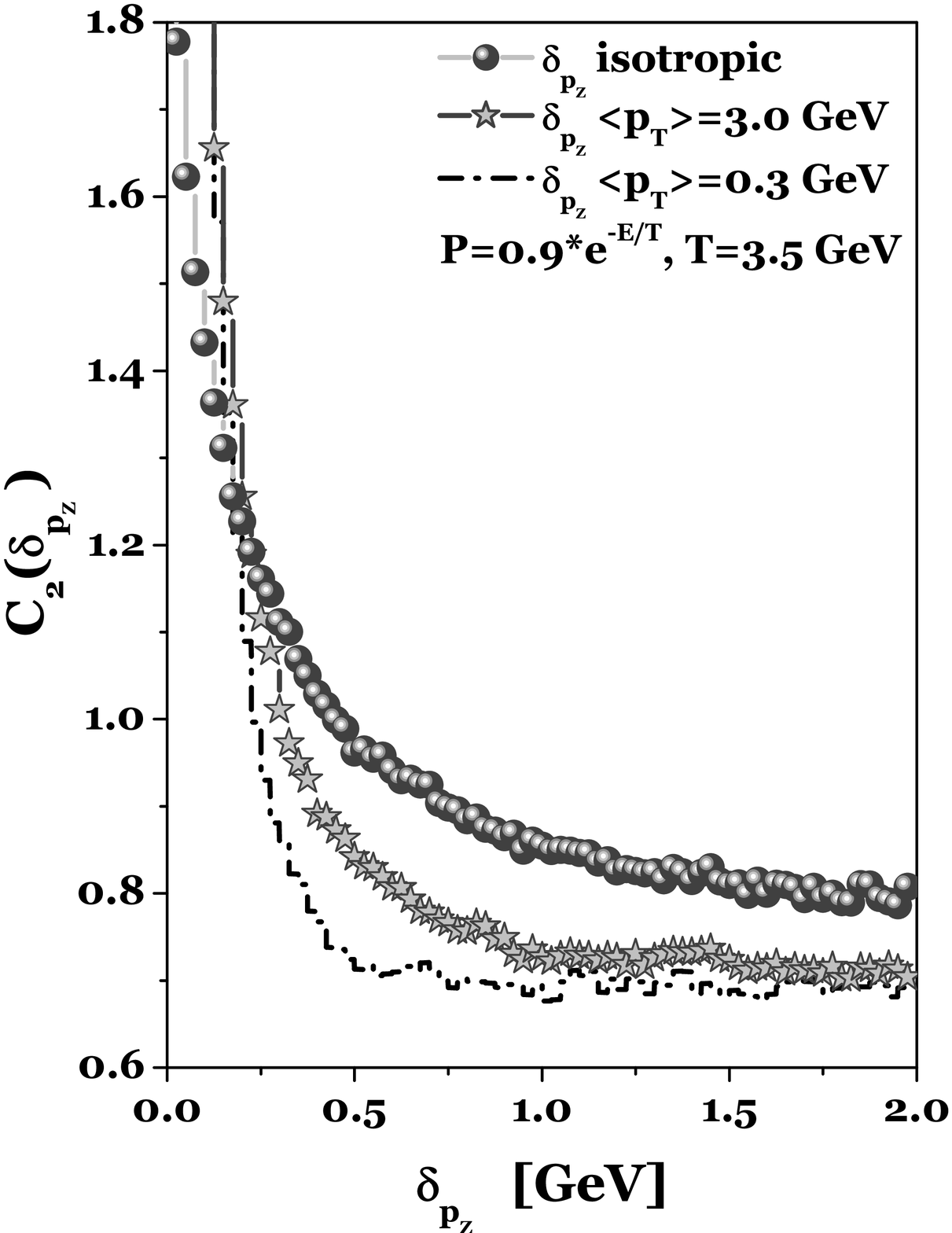}
\end{center}
\vspace{-0.6cm} \caption{Comparison of two different ways of
choosing momenta of particles occupying given EEC ($|\vec{p}|$ is
always fixed) for $\delta_{p_x}$ (left panels) and  $\delta_{p_z}$
(right panels).  Upper panels:  the $(+,-)$ sign of $p_L $ was
chosen randomly for every particle without referring to EEC it
belongs to.  Lower panels: it was chosen randomly for EECs and kept
the same for all particles in it. In all cases three choices of
$p_T$ is shown: unlimited ({\it isotropic}) and with limits imposed
by two different values of $\langle p_T\rangle$ when sampling $p_T$
from exponential distribution; in all cases $\sigma_0 =0$ .}
\label{fig:SvsE} \vspace{-0.3cm}
\end{figure}
\begin{figure}[!htb]
\begin{center}
\includegraphics*[width=4.cm]{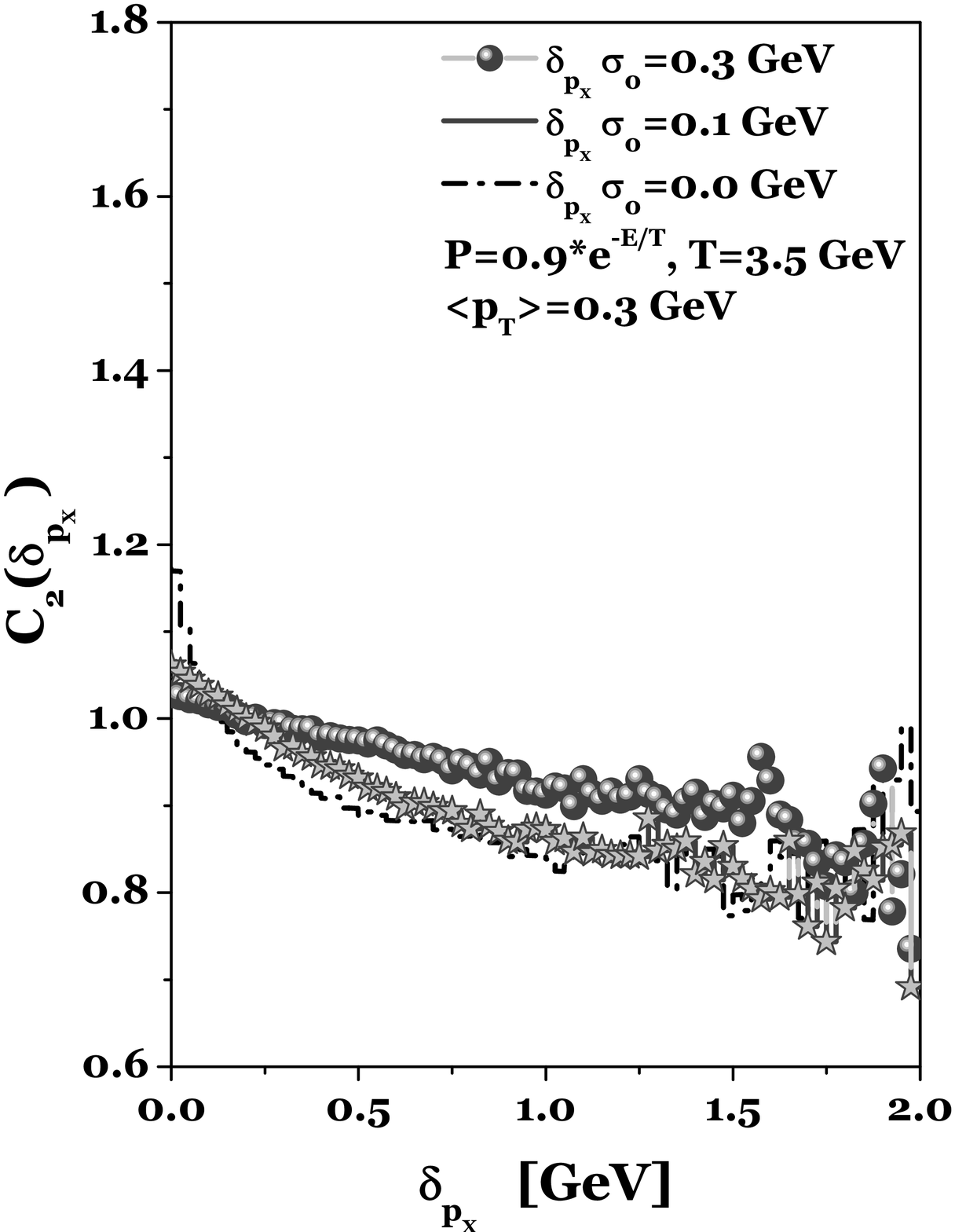}
\includegraphics*[width=4.cm]{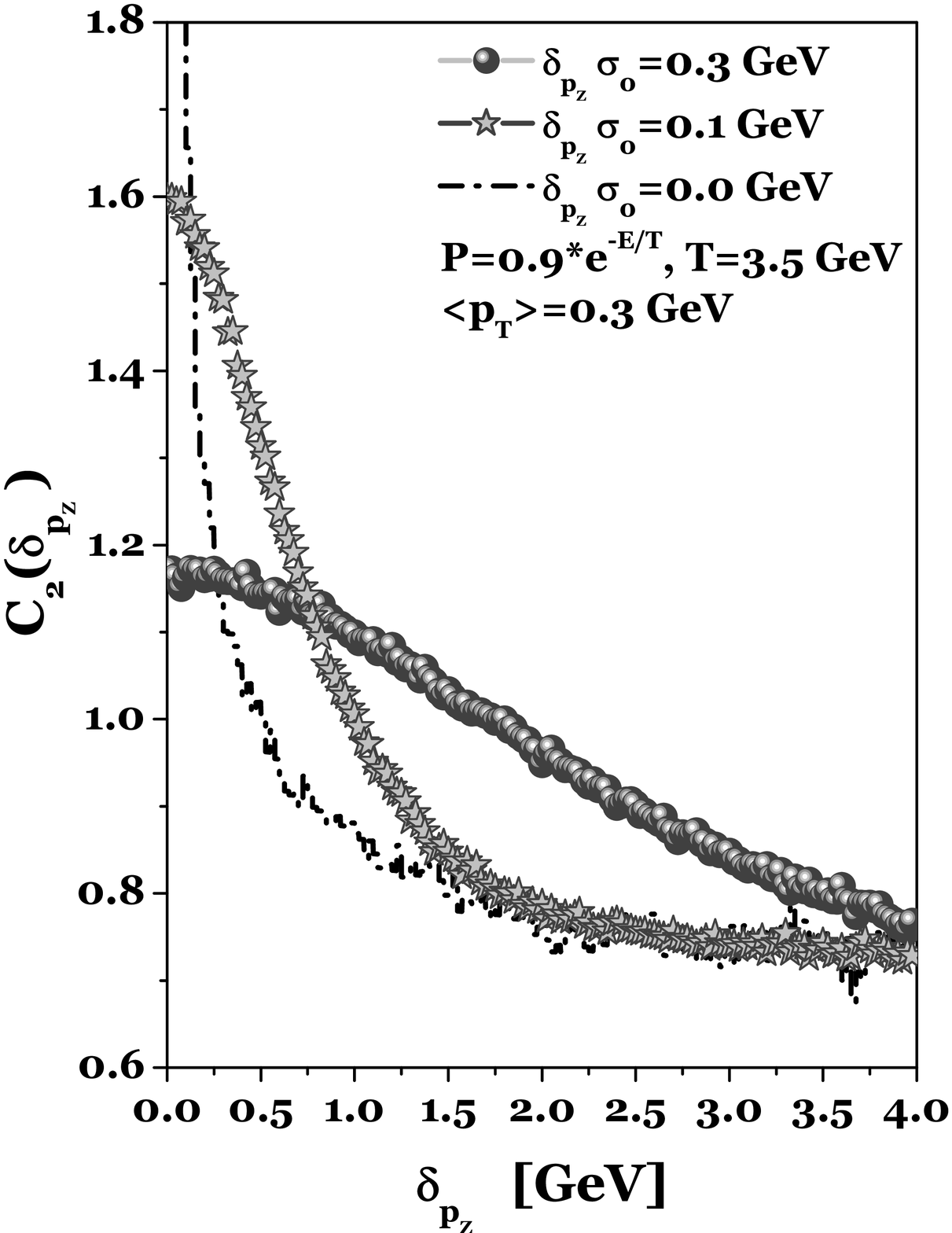}\\
\includegraphics*[width=4.cm]{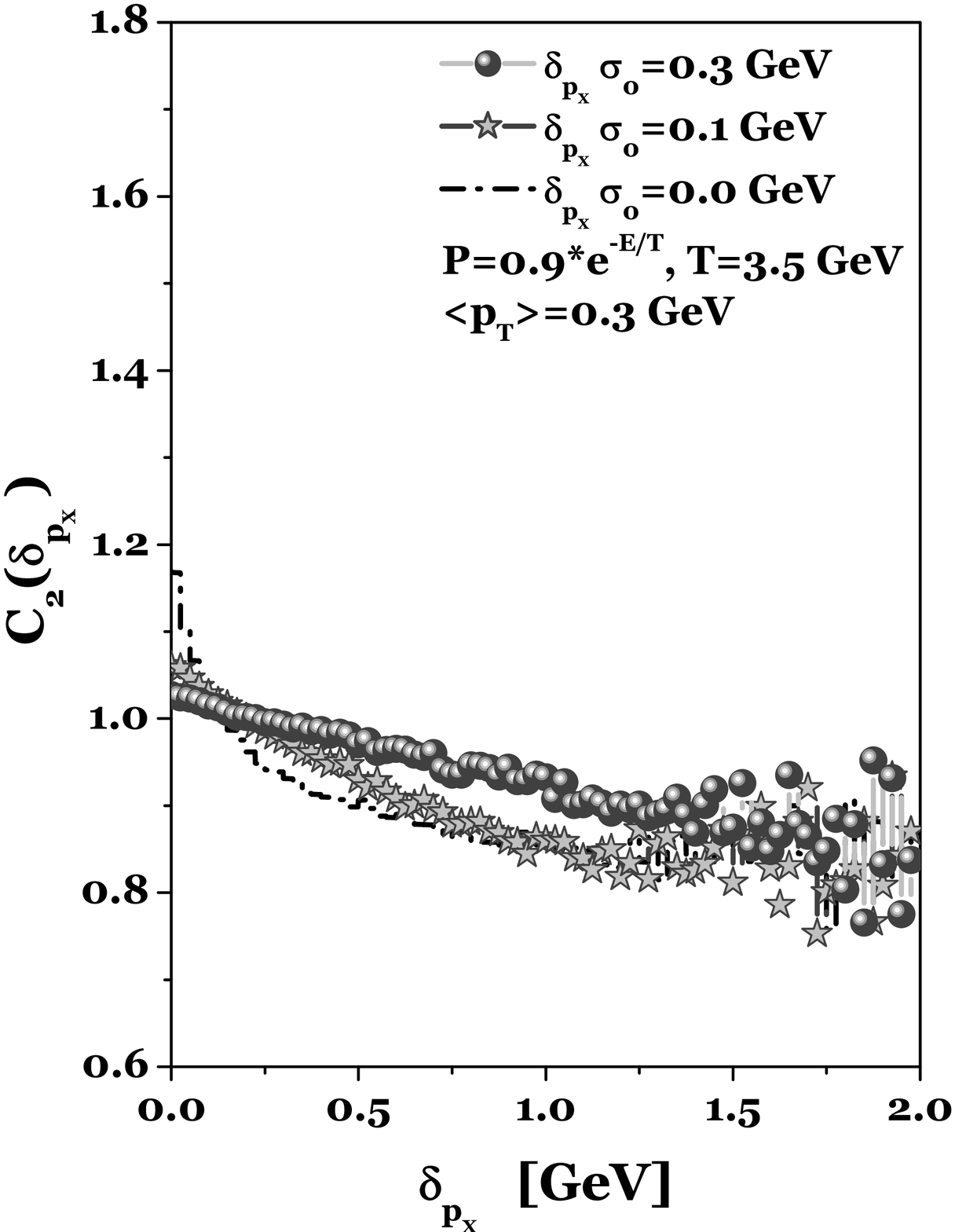}
\includegraphics*[width=4.cm]{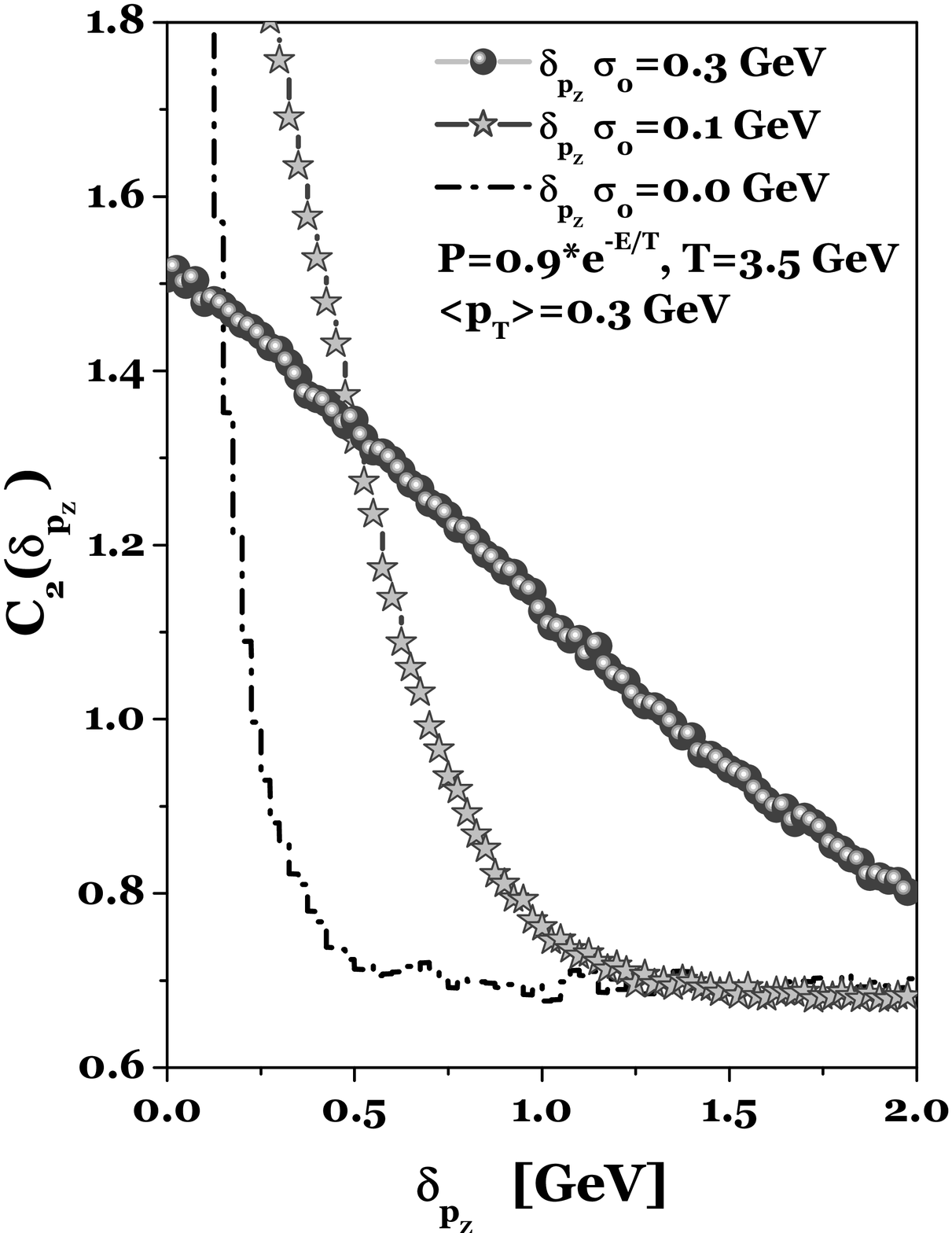}
\end{center}
\vspace{-0.6cm} \caption{The same as in Fig. \ref{fig:SvsE} but for
choice of $p_T$ restricted by $\langle p_T\rangle = 0.3$ GeV and for
different values of $\sigma_0 = 0.0$, $0.1$ and $0.3$.}
\label{fig:SvsEsdep} \vspace{-0.3cm}
\end{figure}

If one wants to further continue with  $C_2 \left(
\delta_{p_{x,y,z}} \right)$, some additional input (new
parameter(s)) is necessary, which has to be justified. We shall not
discuss this point in detail, as it would bring us outside the main
scope of this paper. Instead, we limit ourselves to showing in Fig.
\ref{fig:SvsE} the results of some more refined choices of
directions of momenta. All results are for $\sigma_0=0$,
introduction of nonzero $\sigma_0$ will change them accordingly in
the manner presented in Fig. \ref{fig:Kozlov}. They were obtained by
choosing first values of transverse momenta, $p_T=|\vec{p}_T|$, by
selecting them from some exponential distribution constrained only
by the assumed mean value, $\langle p_T\rangle$, which serves
therefore as new parameter. This corresponds to selection of polar
angles from the band centered on $\Theta_{mean} =
\arctan\left(\langle p_T\rangle /p_L^{(max)}\right) $), the
corresponding axial angles were chosen uniformly from the $[0,2\pi
]$ range. In this way, one gets components $p_x$ and $p_y$ and
longitudinal component $p_z = p_L = \pm \sqrt{p^2 - p^2_T}$. Two
natural situations are considered here: $(i)$ - the maximally
isotropic case (the $p_L$ of every consecutive particle,
irrespectively of the EEC they belong to, is randomly assigned the
sign $(\pm)$) and $(ii)$ - the case in which bunching in energies is
preserved also on the level of momenta (all particles in a given EEC
have $p_L$ in the same hemisphere). All other choices should
interpolate between these two.

Two characteristic features seen in Fig. \ref{fig:SvsE} should be
noticed: $(i)$ one observes narrowing of $C_2(\delta_{p_x})$ (i.e.,
transverse) distributions with tightening the allowed $p_T$ region
(i.e., when proceeding from fully isotropic distributions to those
restricted by assumed $\langle p_T\rangle$ with diminishing values
of $\langle p_T\rangle$), this effect is essentially independent of
the way the signs of $p_z$ components are chosen. $(ii)$
$C_2(\delta_{p_z})$ shows different behavior depending on which
choice of signs is followed: it shows no dependence on $\langle
p_T\rangle$ for the choice $(i)$ above whereas for the choice $(ii)$
a difference shows up only for $\delta_{p_z} > 0.2$ GeV. The
situation changes dramatically when one allows for smearing of
energy in the EEC, i.e., for $\sigma_0 > 0$, see Fig.
\ref{fig:SvsEsdep}. This means introducing a new parameter, to which
our results are most sensitive but which is still not well
understood (see, for example Appendix \ref{sec:AA}).

\begin{figure}[!htb]
\begin{center}
\includegraphics*[width=4.0cm]{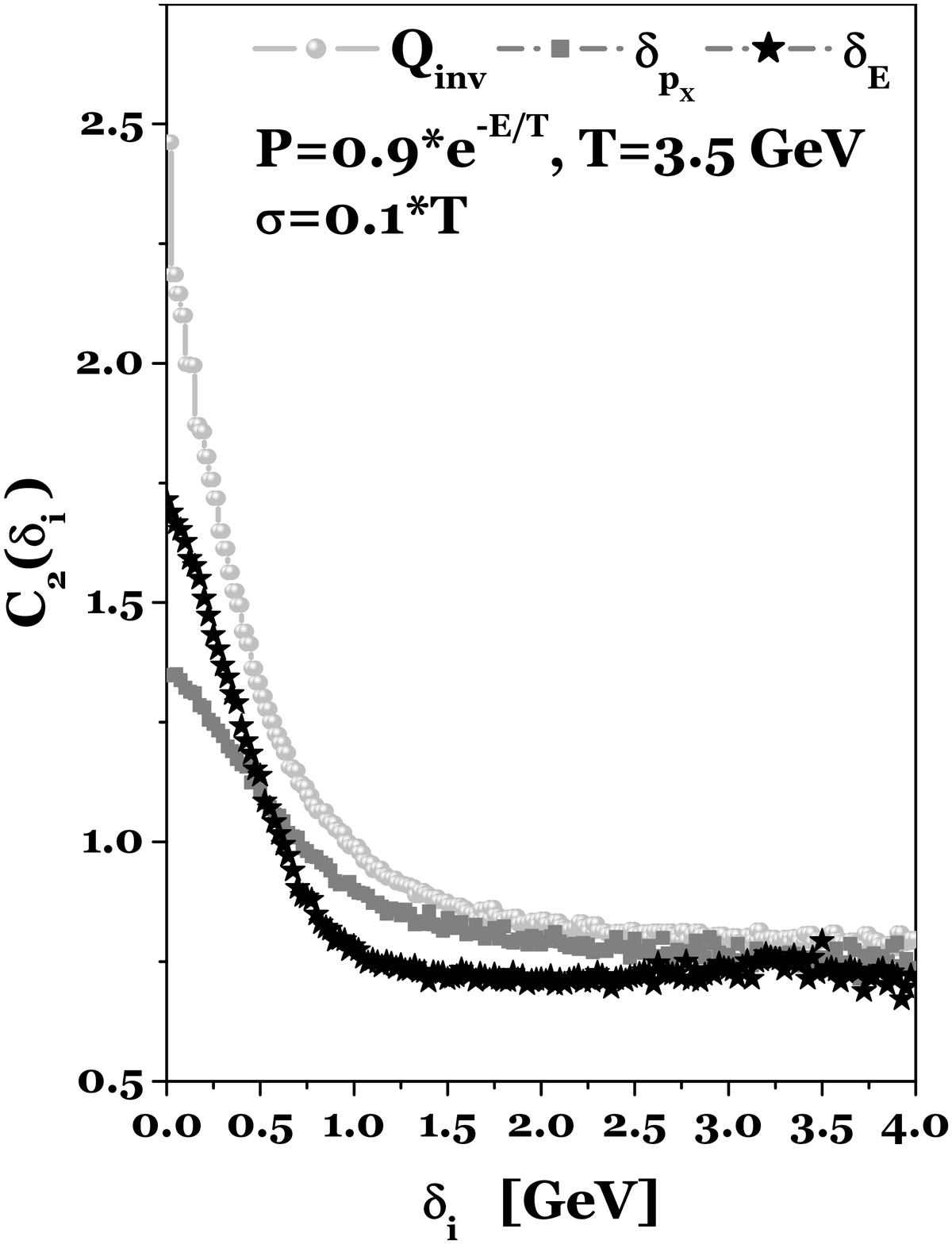}
\includegraphics*[width=4.0cm]{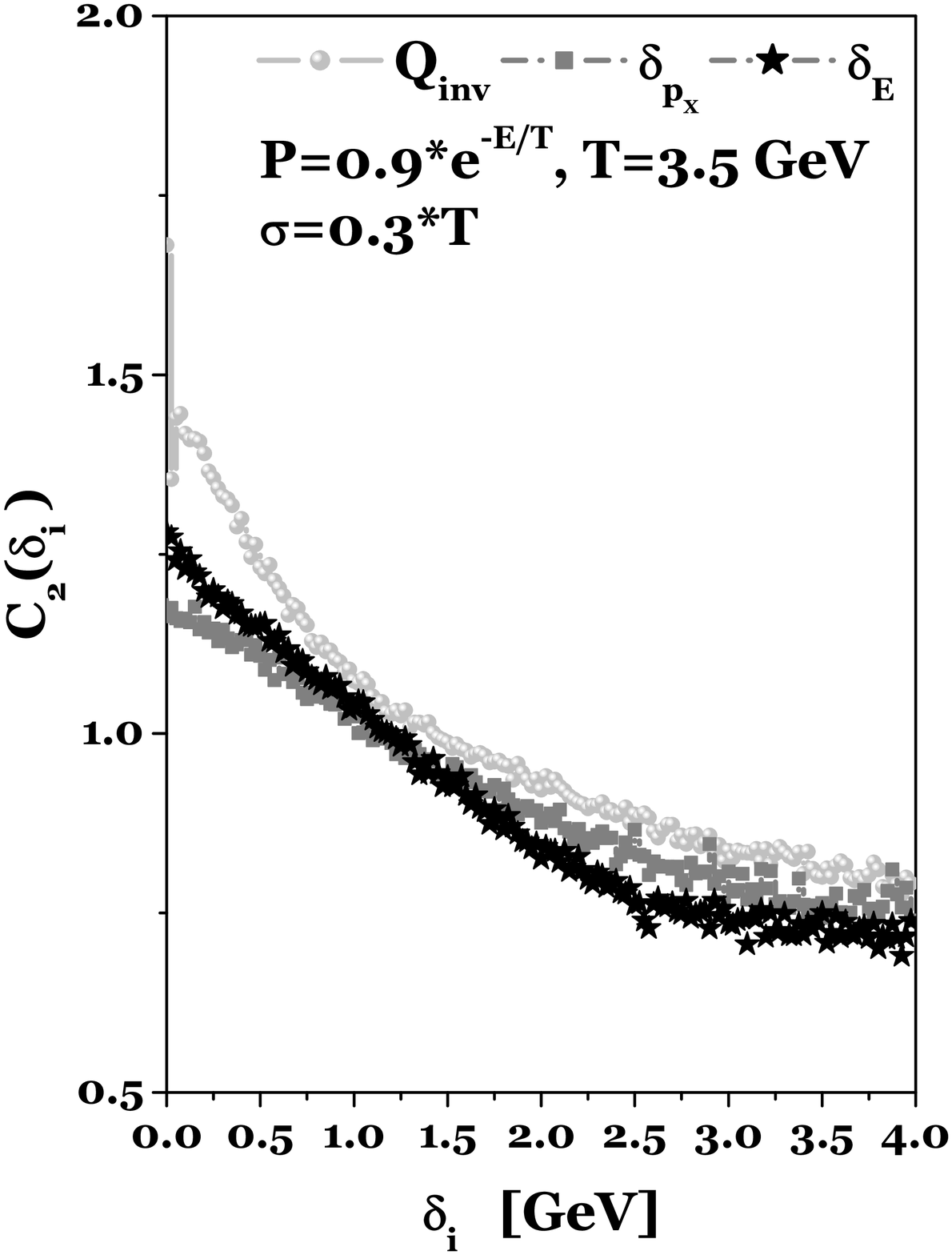}
\end{center}
\vspace{-0.6cm} \caption{Comparison of
$C_2(\delta_i=\delta_{E,p_x})$ as presented in Fig. \ref{fig:Kozlov}
for $\sigma_0=0.1$ with $C_2(\delta_i=Q_{inv})$ for the same
parameters (left panel). Notice that maximum of $C_2(Q_{inv}=0)> 2$
and to get it below this value one has to increase $\sigma_0$ (see
right panel with $\sigma_0 = 0.3$).} \label{Qinv} \vspace{-0.3cm}
\end{figure}
In Fig. \ref{Qinv} results for $C_2(\delta_E)$ and
$C_2(\delta_{p_x})$ are compared with those for $C_2(Q_{inv})$,
where $Q^2_{inv} = - \left(p_1 - p_2\right)^2 $ with $p_{1,2}$ being
$4$-momenta of particles $1$ and $2$ (it is relativistically
invariant variable , such that $Q_{inv}^2 =0$ implies for massive
particles that $\vec{p}_1=\vec{p}_2$ and BEC has its maximum there).
Notice the peaked shape of $C_2(Q_{inv})$ and the fact that
$C_2(Q_{inv}=0) > 2$ here. As shown in \cite{Q2} both features are
mostly due to the specific kinematics of the $Q_{inv}$ variable
(because of which it collects in the first bin contributions from
the whole range of momenta provided only that they are near enough
to each other). The point is that in case  of $Q_{inv}$ variable
smearing $\sigma$ in energy $E$ and smearing in momentum $p$ are not
independent and therefore they do not sum up (as is the case for
independent variables) but rather they tend to cancel. In fact,
their global effect is of the order of $\sigma\cdot (\sqrt{p^2 +
m^2} - p)$ which tends to zero for large momenta and never exceeds
$\sigma \cdot m$, where $m$ is pion mass. In other words, smearing
in $Q_{inv}$ is considerably smaller then that in energy and
momentum. Therefore there is a natural tendency of increasing
occupancy of the lowest bins in $\delta_i = Q_{inv}$ in comparison
to $\delta_E$ or $\delta_{p_x}$.

\begin{figure}[!htb]
\begin{center}
\includegraphics*[width=7.5cm]{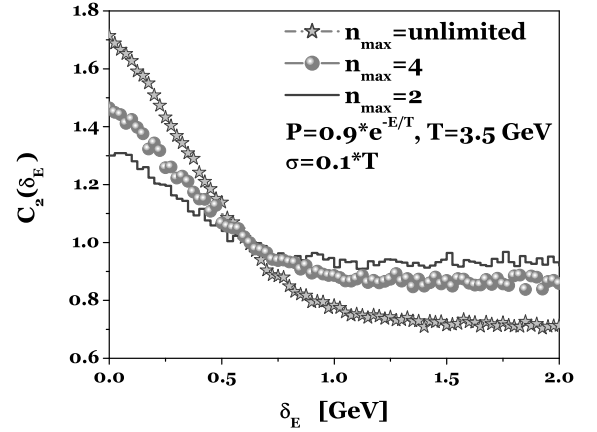}
\end{center}
\vspace{-0.6cm} \caption{Comparison of $C_2(\delta E)$ for different
maximally allowed sizes of EEC given by maximal number of particles
$n_{max}$ they can have. Notice difference from the similar results
obtained for  pionic lattice and presented in Fig. \ref{fig:net}c.
It is caused by the fact that here distance between particles is not
limited by the fixed spacing of the lattice as before.}
\label{multlim} \vspace{-0.3cm}
\end{figure}

\begin{figure}[!htb]
\begin{center}
\includegraphics*[width=4.cm]{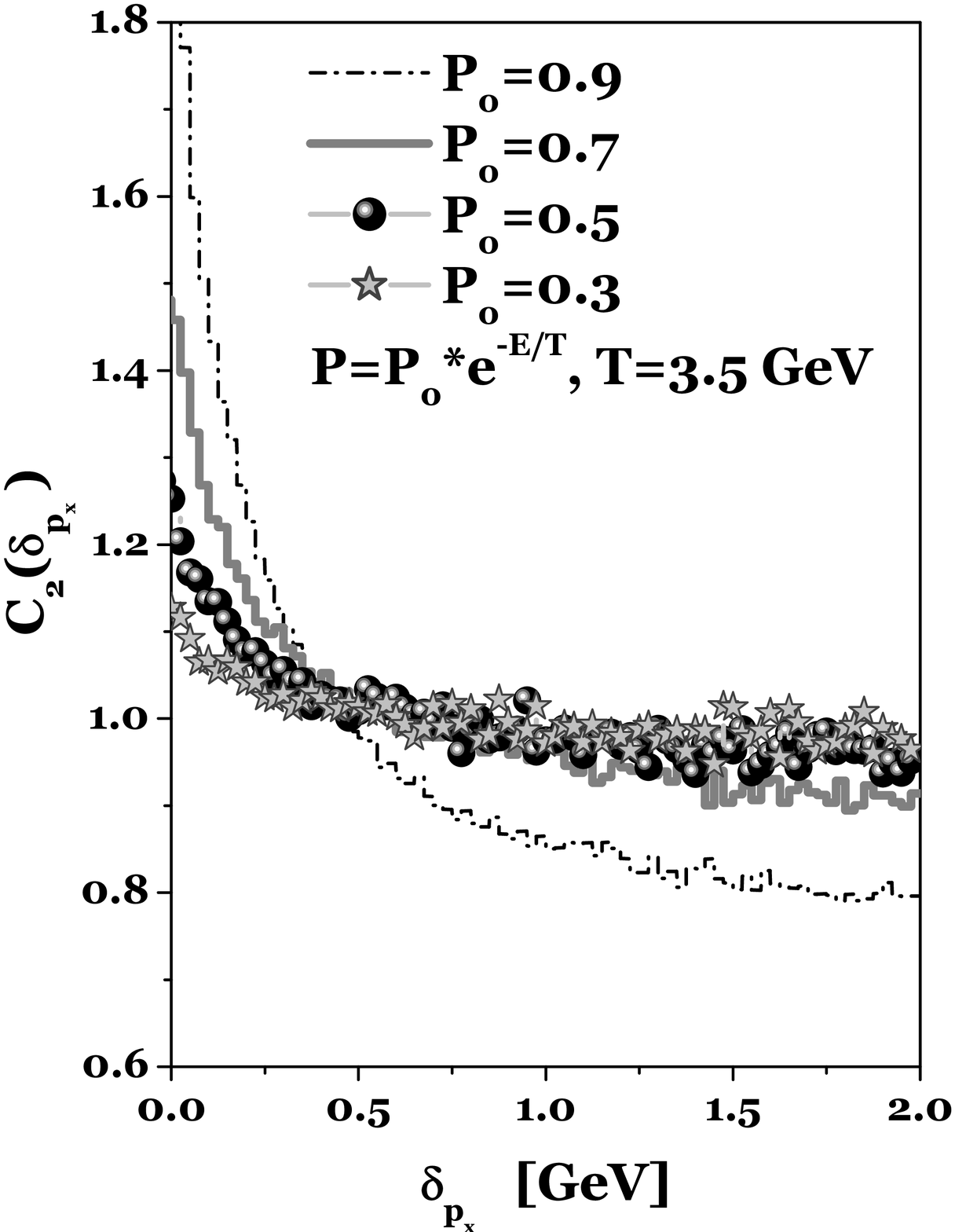}
\includegraphics*[width=4.cm]{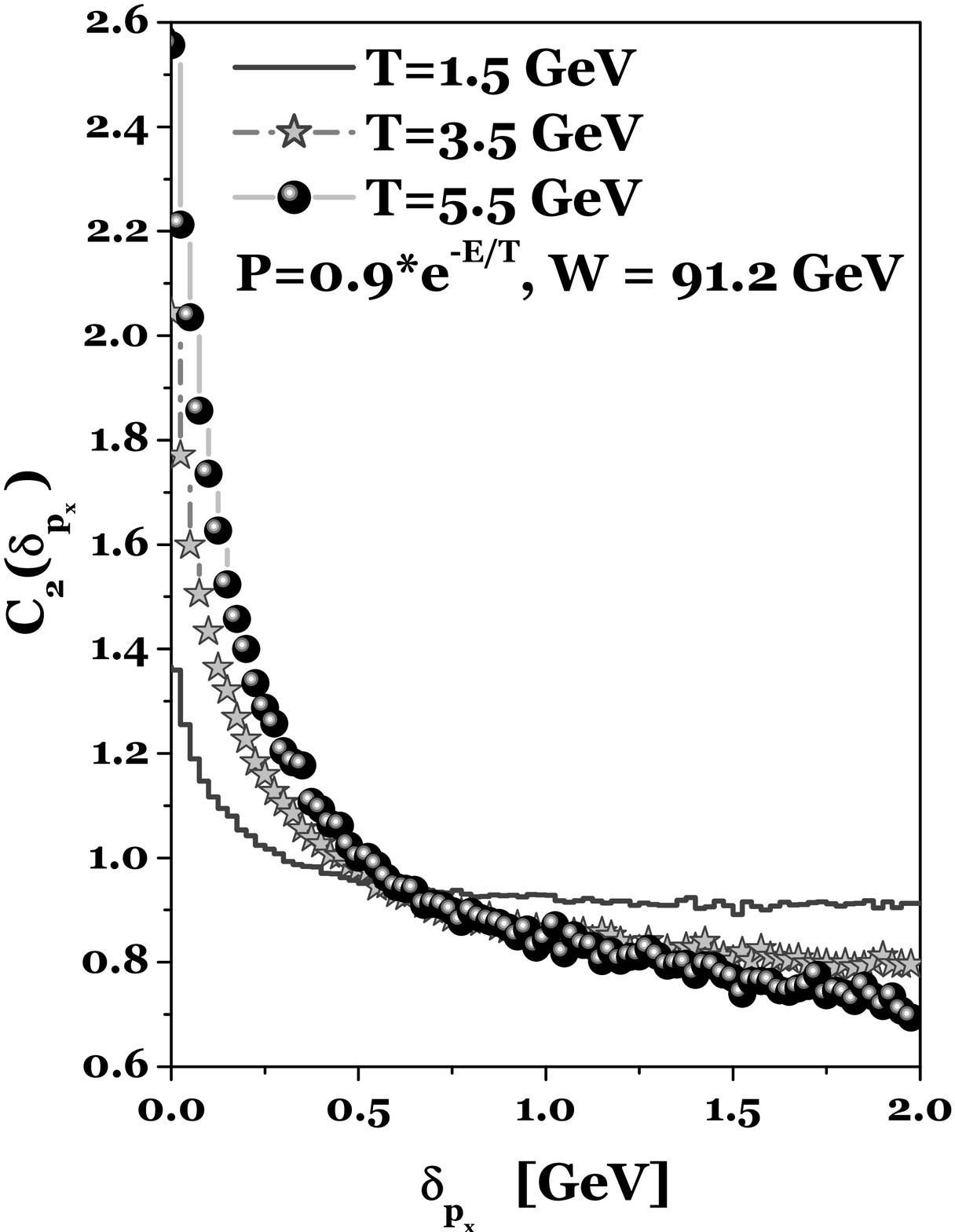}
\includegraphics*[width=4.cm]{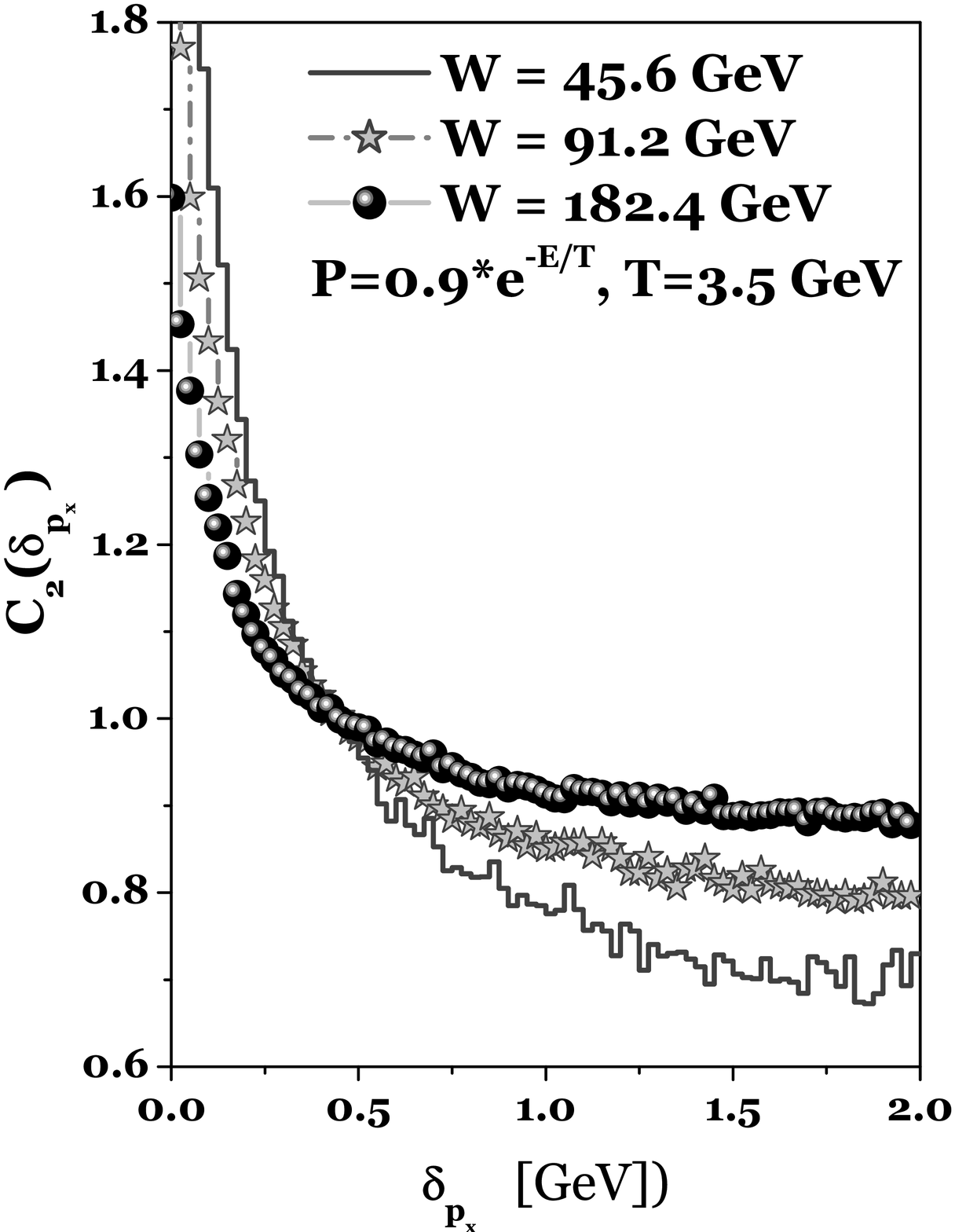}
\end{center}
\vspace{-0.6cm} \caption{Upper panels: sensitivity of
$C_2(\delta_{p_x})$ to different choices of EECs exemplifying by
different sets of parameters used. Lower panel: the same but for
different energies $W$ and fixed size of EEC (as indicated, in all
cases $\sigma_0 = 0$ and isotropic source was considered). The
corresponding values of mean multiplicities, their dispersion as
well as the mean number of EECs and their mean occupancies are
listed in Table \ref{Parameters}.} \label{fig:difpar}
\vspace{-0.3cm}
\end{figure}

Let us now stress the most specific feature of our algorithm: it
{\it always} produces BEC of the highest possible order $n$ (for
which $C_2(\delta_E=0) = n!$). This order is dictated by the the
number of particles in the most populated EEC, which in turn depends
strongly on location of a given EEC in phase space, see Table
\ref{Ecells}. It will significantly influence both the strength (as
given by $C_2(\delta=0)$) and shape of the BEC effect, cf.,  Fig.
\ref{multlim}. In it we show what happens when the maximal number of
particles in each EEC is artificially limited so as not to exceed
some imposed maximal value equal to $n_{max}$. Notice that this
requirement also affects the resulting number of EECs (the smaller
population of EECs the more of them must be present). It is clear
that this effect will be strongest in events with very high
multiplicities recorded (cf. \cite{LargeN} for references to
projects of the respective experiments). The sensitivity of our
algorithm to $n_{max}$'s presented in Fig. \ref{multlim} (for
$C_2(\delta_{p_x})$ the changes are qualitatively the same for all
choices of momenta presented here) makes it an ideal tool for
numerical investigations of BEC also for particles satisfying
statistics different than BE (for example, the so called {\it
parastatistics} \cite{parabosons}).

We close this Section showing how sensitive are $C_2(\delta_i)$ to
different choices of EECs represented by different values of
parameters $\mathcal{P}_0$ and $T$ and to different masses $W$ of
hadronizing source, cf. Fig. \ref{fig:difpar} and Table
\ref{Parameters} (this is done again for $\sigma_0=0$ and using
isotropic distributions of directions of momenta. Any changes in
them, as discussed before, would then change these results
accordingly in the way demonstrated in Figs. \ref{fig:Kozlov},
\ref{fig:SvsEsdep} and \ref{fig:SvsE}). The most characteristic
feature is the observed growth of $C_2(\delta_i=0)$ (i.e., the grow
of the so called "parameter of chaoticity" $\lambda =
C_2(\delta_i=0)-1$ ) with diminishing number of EECs, $\langle
N_{cell}\rangle$. Actually, this result was behind the original
introduction of the notion of EECs in describing the BEC effect done
in \cite{BSWW} where the number of EECs is  tightly connected with
the parameter $k$ in the NB multiparticle distributions used to fit
data. Interesting feature of this approach, shared by our picture as
well, is that, as shown in \cite{BSWW}, it explains in a natural way
the dependence of $\lambda$ on $dN/dy$ and on the atomic number of
projectiles $A$ \cite{FOOThalo}. In Appendix \ref{sec:CC} we discuss
changes in $C_2(\delta_i)$ introduced when correcting for the
inevitable energy-momentum and charge imbalances induced by our
algorithm.

\subsection{\label{sec:Discussion}Discussion}

Let us recapitulate the physical picture we are proposing. Its basic
object is an EEC, a quantum state containing a number of identical
secondaries of the same, or nearly the same energies. These
secondaries are assumed to satisfy the BE statistics which is
imposed by demanding that they follow geometric distribution. As a
result the correlation functions $C_2(\delta_X)$ that follow are
very sensitive to the characteristics of the EEC, for example the
width of $C_2(\delta_X)$ is proportional to the allowed energy
spread in EEC, $\sigma$. It is best seen on the example of
$C_2(\delta_E)$, cf. Fig. \ref{fig:Kozlov}. Interestingly enough,
when all particles in the EEC have the same energy, $C_2(\delta_E)$,
is divergent. On the other hand, in the same situation, correlation
functions in momenta have already  nonzero widths. This is because
the choice $\sigma=0$ does not constraint directions of momenta. The
simplest case of isotropically selected directions, represented by
$C_2(\delta_{p_x})$, is shown in Fig. \ref{fig:Kozlov}. The choice
of directions provides therefore additional freedom in modeling
$C_2(\delta_{p_{x,z}})$. It can be seen in Figs. \ref{fig:SvsE} and
\ref{fig:SvsEsdep} where examples of restricting the range of
transverse momenta and the choice of longitudinal momenta are
displayed. As for the physical meaning of $\sigma >0$, it is
tempting to identify it with the temporal characteristic of
hadronization process, in fact with the life time of an average EEC.

So far we presented results only for direct pions being produced. To
include resonances one would first have to decide whether the BEC
should affect them in the same way as particles or whether they
affect only the pions resonances decay into. This point surely
deserves further discussion, but it would bring us outside the scope
of our paper. Therefore we present in Fig. \ref{resonances} results
for $C_2$ calculated for $\rho$ mesons (of charges $(+,-,0)$ and
mass $m_{\rho} = 0.7$ GeV, with zero widths assumed for simplicity)
considered to be simple particles subjected to the same procedure of
building EEC as that for pions. This is compared with the case where
such  $\rho$'s subsequently decay into pions.
\begin{figure}[!htb]
\begin{center}
\includegraphics*[width=6cm]{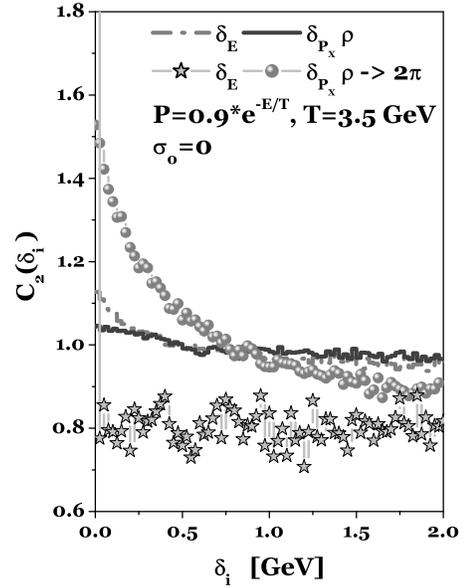}
\end{center}
\vspace{-0.6cm} \caption{Comparison of BEC for mesons $\rho$ (with
$m_{\rho}=0.7$ GeV) and for pions obtained from their decays. To
maximize effect we use the EEC's for $\rho$'s with fixed energies
(i.e., $\sigma_0 = 0$). Both $C_2(\delta_E)$ and $C_2(\delta_{p_x})$
are presented. Notice that also for $\rho$ correlation function
$C_2(\delta_E)$ is concentrated in first bin only.}
\label{resonances} \vspace{-0.3cm}
\end{figure}
Notice that whereas the BEC for $\rho$'s is quite strong and not
very much different from that for pions only, it hardly survives the
process of $\rho$'s decay, especially for the $C_2(\delta_{p_x})$
case. However, it must be stressed at this point that, so far, our
$\rho$'s were treated as spinless particles and that they were
assumed to decay into pairs of pions in an isotropic way in their
center of mass \cite{FOOT12}.

In our algorithm only particles in EEC are subjected to BEC, there
are no intercorrelations of the BE type between particles from
different EECs (see Appendix \ref{sec:A}). Such picture seems to be
supported by recent data on BEC in $e^+e^- \rightarrow W^+W^-$
multi-$W$ boson production which show non existence of inter-$W$
BEC. This result suggest strongly that although spatially located
practically on top of each other, nevertheless bosons $W$ act as
{\it independent} sources of pions in this case \cite{eeWW}.

We would like to stress here that all restrictions on energies and
momenta mentioned above influence first of all the shape of a single
EEC in momentum space rather than global characteristic of a given
hadronizing source used. It is therefore to a large extent the EEC
information on which are encoded in the correlation function $C_2$.
On the other hand their number depends in the first instance on the
characteristics of the hadronizing source encoded in the choice of
$f(E)$. Occupancy of EEC, however, depends strongly on
$\mathcal{P}_0$ and together with $T$ from $f(E)$ change
substantially both the multiplicity distributions (in our case from
the Poisson-like to P\'{o}lya-Aeppli one) and single particle
distributions as well (see Figs. \ref{fig:cells} and
\ref{fig:Edist_s0}).

\begin{figure}[!htb]
\begin{center}
\includegraphics*[width=4cm]{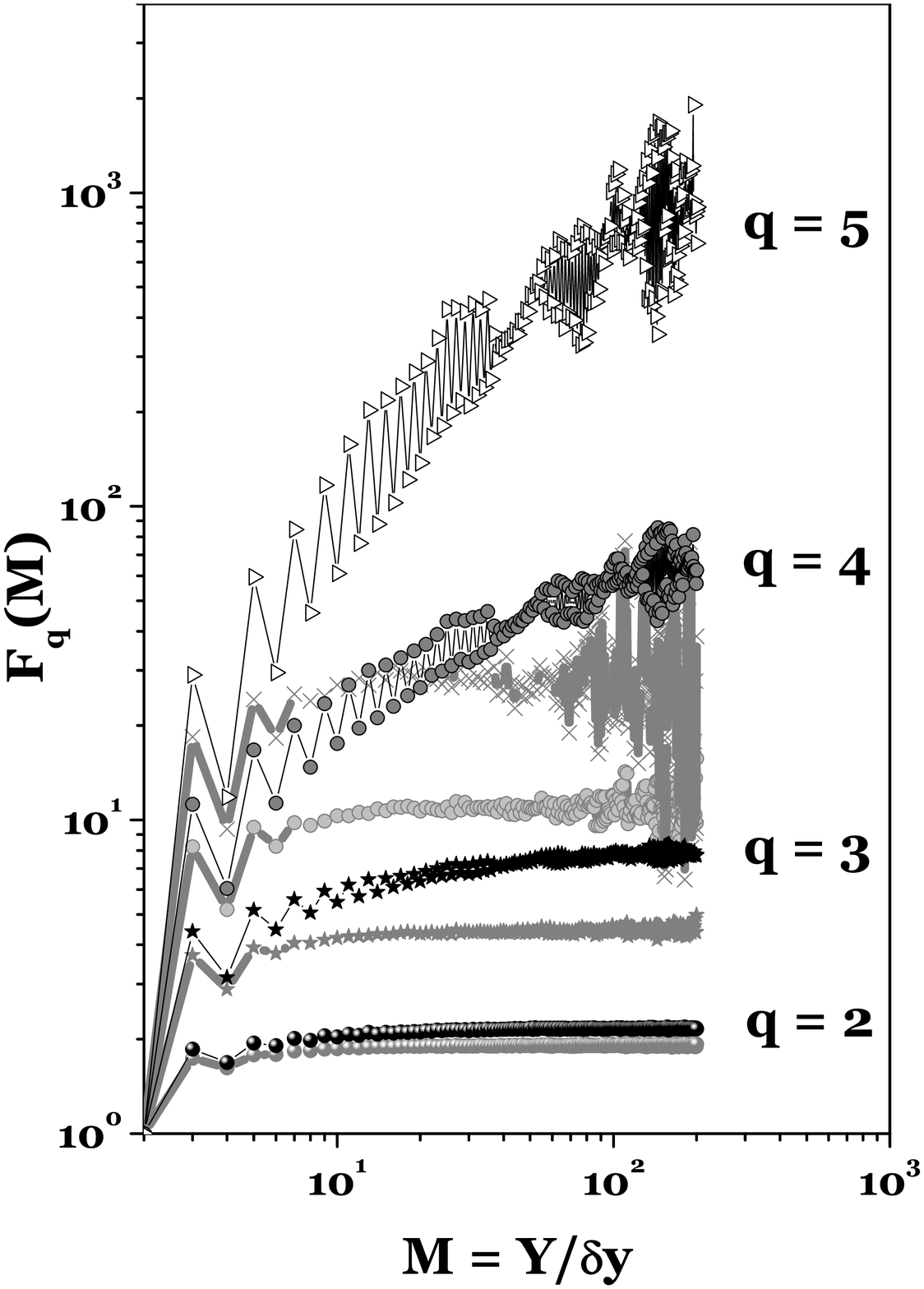}
\includegraphics*[width=4cm]{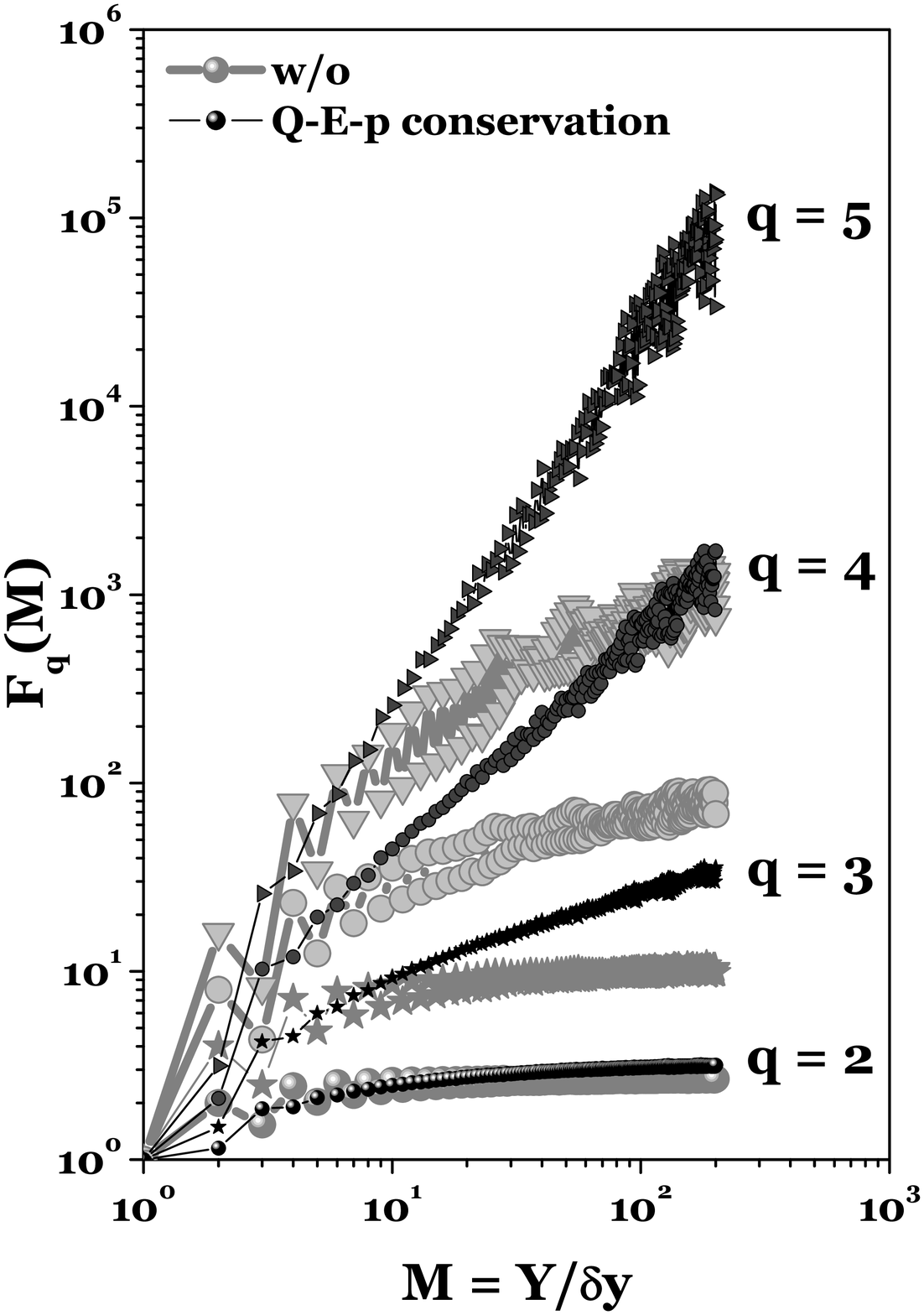}
\end{center}
\vspace{-0.6cm} \caption{Example of intermittency signals obtained
using our algorithm with $\mathcal{P}_0=0.9$, $T=3.5$. Left panel:
not corrected for energy-momentum and charge imbalances results for
$\sigma_0=0$ (black curves) and $\sigma_0=0.3$ (grey curves) are
compared. Right panel: results for $\sigma_0=0$ only with and
without corrections (most of the visible effect comes from
correcting the energy momentum imbalance). Moments $F_q(M)$ are
defined as in \cite{Intermitency}.} \label{Inter} \vspace{-0.3cm}
\end{figure}

It is worth to mention at this point that together with BEC our
algorithm introduces also some intermittency signal, much in the way
expected already in \cite{FOOT6}, see Fig. \ref{Inter}. It depends
noticeably on the smearing of EEC in energy and is very sensitive to
the energy momentum imbalance corrections (cf., Appendix
\ref{sec:CC}) \cite{Sarki}.

\section{\label{sec:summary}Summary and conclusions}

To summarize:  we argue that proper numerical symmetrization of the
multiparticle state of identical particles can be achieved in an
economical way (in what concerns computational time) only by
bunching them in phase space in such way that (identical) particles
in each bunch (called here EEC - {\it elementary emitting cell})
have (almost) the same energies and follow a geometrical
(Bose-Einstein) distribution. Only particles in EEC experience BEC,
those from different cells do not (see Appendix \ref{sec:A}). We
regard this conjecture as emerging in a natural way from previous
investigations \cite{zajc,MP,Cramer,ITM,AIP}. It is the main result
of studies of a properly symmetrized multiparticle wave function of
identical secondaries produced in the reaction
\cite{zajc,MP,Cramer}. They unravelled that in such state the
originally uniformly distributed particles start to bunch
\cite{zajc,Cramer}, also changing the original poissonian
multiplicity distribution to a NB one \cite{zajc}. The similarity
with the clan model \cite{NBD} leading to QCM proposed here was then
immediate. The same conjecture was achieved independently following
studies in which one works with number of quanta (particles) without
invoking wave functions \cite{Zal,Pur,GN,OPTICS,BIYA,QS,F}. In the
first practical application of the concept of bunching \cite{ITM}
the whole (one dimensional) phase space was divided into such EECs
(of equal size in rapidity space) and this simple decision resulted
in profound consequences for what concerns the ability to describe
different physical distributions. We develop this idea further: our
EECs are formed dynamically, they can both overlap each other and be
widely separated from each other and their number and multiplicities
(i.e., their sizes) of particles in them fluctuate from event to
event. This work is then about how to form such EECs and what to do
with them.

\begin{figure}[!hbt]
\vspace*{0.2cm}
\begin{center}
\includegraphics*[width=7.cm]{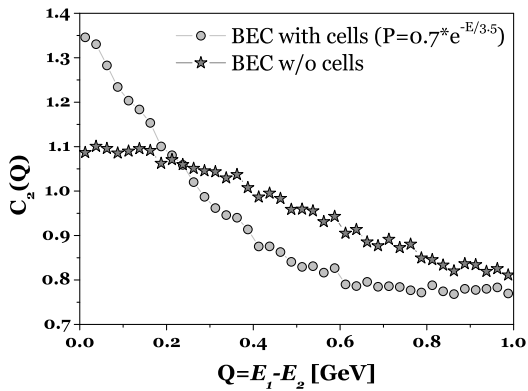}
\end{center}
\caption{Comparison of $C_2$ modelled by using MC event generators
with EEC (circles) and without EEC's but with selecting particles
directly from the corresponding Bose-Einstein distribution $\langle
N(E_i)\rangle$ as given in eq. (\ref{eq:BEE1}) \cite{THERMINATOR}.
In both cases the Boltzmann distribution for classical particles was
used as reference event. This fact may be crucial to get apparent
increase in case of using directly Bose-Einstein distributions (as
in \cite{THERMINATOR}). The use of mixed event instead, which will
not affect substantially the result obtained using EECs, will most
probably kill the other effect.}
\label{Comparison} 
\end{figure}

One can ask whether cells (or proper bunching of identical particles
in phase-space, advocated here) are really needed to model quantum
Bose statistics or, perhaps, is would be enough just to use the
usual Bose-Einstein distributions to this aim. To answer this
question let us compare two ways of producing bosonic particles:
$(i)$ generating them directly from Bose-Einstein distribution (like
$\langle N(E_i)\rangle$ as given in eq. (\ref{eq:BEE1}), which is
the most simple way advocated on many occasions) and, $(ii)$
generating particles from a Boltzmann distribution and bunching them
in an appropriate way in phase-space according to QCM presented
here. In the first case, one gets the correct single particle
distribution, whereas in the second case one  also accounts for
multiparticle correlations of quantum statistical origin. The
corresponding results are presented in Fig.\ref{Comparison}. They
show that both approaches lead to very different results, which can
be understood by realizing that case $(i)$ corresponds to a
particular realization of case $(ii)$, namely with only one cell
containing the same average number of particles. In fact case $(i)$
shows only some trivial correlations which can be eliminated by a
proper choice of the reference distribution. Therefore, we argue
that, according to our understanding, the better approach (after
correcting additionally for charge and energy-momentum
conservations) is that
\begin{equation}
\fbox{\emph{BEC} = \emph{cells} + \emph{geometrical distribution}}
\label{eq:essence}
\end{equation}
From its construction it is evident that our algorithm is best
suited to study events with large multiplicities (as planned in some
experiments \cite{LargeN}). The statement (\ref{eq:essence}) also
summarizes the only possible way to incorporate our algorithm in
MCEG codes: according to our finding it should be done by enforcing
particles to be produced in bunches with characteristics of our
EECs. In fact, closer inspections on all previous efforts to imitate
BEC mentioned in Section \ref{sec:BEC_in_mod} shows that they all
implicitly were aiming in similar direction
\cite{Weights,LUND,Shifts,UWW1,UWW2}. The difference was that they
were trying to do it {\it a posteriori} rather than {\it a priori}
and that their weighting procedure was aimed more at the
spatio-temporal (unknown) characteristics of the hadronizing source
than on the true physical principles of BEC.

In this work we have stopped short of a general comparison with
data. The reason is that we were using  here a most simple
statistical model of hadronization in order to illustrate our
algorithm. According to it a hadronizing source is assumed to be a
single object with some fixed mass $W$ and fixed initial charge,
which is the situation encountered only in $e^+e^-$ annihilation
reactions. In all other multiparticle production processes one
either encounters $W$ varying from event to event (following some
distribution, for example, inelasticity distribution \cite{inel}) or
there is a number of hadronizing sources with different $W$ in each
event (which is the most probable situation in heavy ion
collisions). Moreover, the statistical scheme of hadronization
employed here means that our hadronizing source does not experience
any internal flows and is not subjected to any external force, which
would result in some energy-position correlations or specific
effects of partial coherence (cf., \cite{Kozlov}) and thus
additionally influence the $C_2$ correlation function. The problem
of the possible net charges of such sub-sources was never discussed.
All this asks for very specialized study, which goes outside the
scope of this paper. What we can only say at this moment is that,
when undertaken, such a study would mean the necessity of developing
our formalism further in what concerns details of modeling EECs
mentioned before. The most probable approach would be to {\it
additionally assume} that each EEC itself should be described by a
{\it properly symmetrized} $n_{part}$-particle wave function (where
$n_{part}$ is the number of particles in this EEC). Only then would
one be able to introduce into the model a characteristic
momentum-positions correlations caused by quantum statistics (much
in the spirit leading to eq. (\ref{eq:nwave})). The price for
following such a procedure is the necessity to also introduce to our
description the space-time characteristics of the hadronizing
source, which we were not dealing with so far. This would result in
the same permanent structure as given by eq. (\ref{eq:permanent}),
but with much smaller sizes and with explicit dependence on
spatio-temporal variables (as in eq. (\ref{eq:cut-perm})) to be used
later when selecting momenta (actually, they would be effectively
integrated out in this procedure). However, the noticeable feature
is that in our case this procedure would not demand spatio-temporal
factorization property of the hadronizing source assumed in
(\ref{eq:nwave}). When replacing plane waves used there by Coulomb
distorted wave functions \cite{BCoul} (which is common practice
nowadays) one could then attempt, in principle, to account for the
influence of Coulomb interactions so far neglected here (albeit only
on the level of $2$-body interactions). This is, of course, not the
only possible generalization and therefore we leave the problem of
confrontation with real data for further studies. Finally, let us
notice that any serious comparison with data would have to be done
including corrections for energy-momentum and total charge
imbalances induced by our algorithm, which (as seen in Appendix
\ref{sec:CC}) can be quite substantial and not unique.

Let us close by noticing that, although our algorithm was originally
intended to model quantum phenomena of BEC only, it is in fact more
general. The reason is that the characteristic structure of the
$C_2$ correlation function associated with specific bunching of
identical particles turns out to be quite universal phenomenon also
observed in many other, purely classical systems, provided only that
they exhibit strong and correlated fluctuations \cite{Classical}.
This means then that our algorithm, albeit with different
motivation, could also be applied there. At this point one should
also notice an attempt at numerical modeling of another quantum
phenomenon, namely Bose-Einstein {\it condensation} presented in
\cite{KR} and the possible connection of the above with the physics
of networks \cite{Net}. Finally, we also claim that, because of its
sensitivity to maximal occupancy of EECs, out method
could easily be modified to be able to study BEC effects for para-bosons
\cite{parabosons}.\\

\noindent {\bf Acknowledgements:}  Partial support by the Ministry
of Science and Higher Education (grant 1 P03B 022 30 - OU and GW) is
gratefully acknowledged .\\

\appendix

\section{\label{sec:A}Some remarks on EEC}

We would like to comment in more detail on our proposition that
identical bosons should come in EEC and experience effects of BEC
there, whereas no such effects should be seen when bosons from
different cells are considered. In the language used in
\cite{Glauber} the probability of registration of a coincidence of
$n$ bosons in states $j_1,\dots,j_n$ is given by a normalized
correlation tensor of rank $2n$,
\begin{eqnarray}
g_{j_1\dots j_nj_{n+1}\dots j_{2n}}^{(nn)} &=&
\frac{G^{(nn)}_{j_1\dots
j_nj_{n+1}\dots j_{2n}}}{\prod_{i=1}^{2n}\sqrt{G^{(11)}_{j_ij_i}}}, \quad {\rm where} \label{eq:ggg}\\
G^{(nm)}_{j_1\dots j_n j_{n+1} \dots j_{n+m} } &=& Tr \left\{ \rho
a^{\dag}_{j_1}\dots  a^{\dag}_{j_n} a_{j_{n+1}}\dots a_{j_{n+m}}
\right\}\nonumber
\end{eqnarray}
is given in terms of the density matrix operator $\rho$ whereas
$a^{\dag}$ and $a$ are, respectively, creation and annihilation
operators. If
\begin{equation}
| g^{(nn)}_{j_1\dots j_{2n}}| =1 \qquad  n\leq N
\end{equation}
we have coherence of the order $2N$. Experimentally this means that
the probability of registering $n$ bosons in coincidence is equal to
the product of probabilities to register individual ones. Because of
commutation relations for bosons, $\left[ a_k,a_l \right] = \left[
a_k^{\dag},a_l^{\dag}  \right] = 0$ and $\left[ a_k,a_l^{\dag}
\right] = \delta_{kl}$, one has that $ a_k^{\dag}a_l^{\dag}a_k a_l =
a_l^{\dag}a_k^{\dag}a_k a_l = a_k^{\dag}a_l^{\dag}a_l a_k =
a_l^{\dag}a_k^{\dag}a_l a_k = n_k n_l$ and then from eq.
(\ref{eq:ggg}) that for two bosons from different states
\begin{equation}
g^{(22)}_{k_1l_1k_2l_2} = \frac{\overline{n_kn_l}}{\overline{n_k}\cdot
\overline{n_l}} = \frac{\overline{n_k}\cdot
\overline{n_l}}{\overline{n_k}\cdot \overline{n_l}} = 1. \label{eq:coh}
\end{equation}
This means that two bosons from different states (in our case: from
different EECs) exhibit second order coherene. On the other hand, in
the situation in which only one state $k$ is occupied, i.e., when
$a_k^{\dag}a_k^{\dag}a_k a_k = a_k^{\dag}a_k a_k^{\dag}a_k -
a^{\dag}_k a_k = n^2_k - n_k$, eq. (\ref{eq:ggg}) results in
\begin{equation}
g^{(22)}_{k_1k_1k_2k_2} = \frac{\overline{n^2_k} -
\overline{n_k}}{\overline{n_k}\cdot \overline{n_k}} =
\frac{2\overline{n_k}^2}{\overline{n_k}\cdot \overline{n_k}} = 2
\label{eq:chaos}
\end{equation}
(here we use the fact that in geometrical distribution $ \overline{n_k^2}
- \overline{n_k}^2  = \overline{n_k}(1 + \overline{n_k})$ ). Notice that
it is greater  by unity than (\ref{eq:coh}) but, because  $g$ is not
limited, it cannot serve as degree of coherence. On the other hand we can
write following Sec. \ref{sec:zalewski} the true correlation function
$C_2$ as
\begin{equation}
C_2 = \frac{\overline{n(n - 1)}}{\overline{n}^2} = 1 +
\frac{Var(n)}{\overline{n}^2} - \frac{1}{\overline{n}} ,
\label{eq:CV}
\end{equation}
where $Var(n) = Var(n_k) = \overline{n}(1 + \overline{n})$ for
bosons from the same state, which in our case means from the same
EEC. This immediately leads to $C_2 = 2$ in this case (for Boltzmann
statistics $Var(n) = \overline{n}$ and one gets $C_2 =1$ instead).
For $k$ EECs we have multiplicity distribution in NB form with
variance $Var(n) = kVar(n_k) = \bar{n} + \bar{n}^2/k$ and mean
multiplicity $\bar{n} = k \bar{n}_k$. This leads immediately to $C_2
= 1+ 1/k$, i.e., we have $C_2 =2$ for $k=1$ and $C_2 \rightarrow 1$
for $k \rightarrow \infty$.

\section{\label{sec:AA}Structure of $C_2(Q)$ and finiteness
                       of hadronizing source}

As demonstrated in \cite{Kozlov} using a quantum field theory
approach to BEC, the structure of the correlation function $C_2(Q)$
is connected with the finiteness  of the hadronizing source. In the
four-dimensional case, the wave function formalism with $\exp \left(
-i p_{\mu}\cdot x^{\mu}\right)$ implies the infinite
($4$-dimensional) volume of the hadronizing source, $V\rightarrow
\infty$. This can then be connected to the fact that commutation
relations for the respective operators contain delta functions,
\begin{equation}
\left[\hat{c}(p_{\mu}),\hat{c}(p'_{\nu}) \right] =
\delta^4(p_{\mu}-p'_{\nu}),
\end{equation}
which are nonzero only for identical values of the four-momenta (i.e.,
also energies), . They in turn lead to correlation functions $C_2(Q)$ in
the form of
\begin{equation}
C_2(Q) = 1 + \delta(Q\cdot R),
\end{equation}
i.e., to $C_2(Q) > 1$ but only at one point, $Q=0$ (cf., Table
\ref{Kozlov}).
\begin{table}[h]
\caption{Schematic presentation of how a different form of
commutation relations in a quantum field theoretical description of
BEC results in a different form of $C_2$ function \cite{Kozlov}. The
actual form of $C_2$ (i.e., of the function $\mathfrak{g}(Q\cdot
R)$) depends on the form of function $\Delta(\dots)$ used to
moderate the original $\delta(\dots)$ function.}
\begin{center}
\begin{tabular}{||c|c|c|c||}
\hline
 & & & \\
Volume  & Wave function & Commutation &
Correlation \\
($4dim$) & & relation  &  function \\
& & $\left[\hat{c}(p_{\mu}),\hat{c}(p'_{\nu}) \right] $ & $C_2(Q)-1 $ \\
 & & & \\
\hline
 & & & \\
~$V\rightarrow \infty$~ & ~$e^{-i p_{\mu}\cdot x^{\mu}}$~ & ~$
\delta^4(p_{\mu}-p'_{\nu})$~ &
~$ \delta(Q\cdot R)$~  \\
 & & & \\
\hline
 & & & \\
~$V=V_0$~ & ~$e^{-i p_{\mu}\cdot x^{\mu}-\frac{1}{2\sigma_p^2}p^2}$~
& ~$ \propto \Delta^4(p_{\mu}-p'_{\nu})$~ &
~$\mathfrak{g}(Q\cdot R)$~  \\
 & & & \\
\hline
\end{tabular}
\end{center}
\label{Kozlov}
\end{table}
However, as shown in \cite{Kozlov}, assuming commutation relations
in the form of some sharply piked (but not infinite) functions
$\Delta$, i.e., replacing delta functions  by functions with
supports larger than limited to a one point only,
\begin{equation}
\left[\hat{c}(p_{\mu}),\hat{c}(p'_{\nu}) \right] \propto
\Delta^4(p_{\mu}-p'_{\nu}),
\end{equation}
results immediately in the correlation function endowed with final width:
\begin{equation}
C_2(Q) = 1 + \mathfrak{g}(Q\cdot R)
\end{equation}
where the form of $\mathfrak{g}(Q\cdot R$ is a straightforward
reflection of the assumed form of $\Delta^4(p_{\mu}-p'_{\nu})$).
Such a procedure corresponds to introducing a finite dimension of
the source, $V=V_0$ (and the use of wave packets formalism,
$\exp\left[ -i p_{\mu}\cdot x^{\mu} - p^2/\left(2\sigma_p^2\right)
\right]$, instead of plane waves).

\section{\label{sec:CC}Correcting for energy-momentum and charge imbalances}

Any random selection procedure, even when following formulas which
assume exact energy-momentum and charge conservations, frequently
induces some energy-momentum and charge imbalances and this is also
true in the case of our algorithm \cite{correl-cons}. The obvious
remedy would be to only accept events preserving the initial values
of energy-momentum and charge. This, however, would result in an
unacceptable long computational time. To the best of our knowledge,
only in the algorithm presented in Section \ref{sec:japan}
\cite{ITM} is energy-momentum assured using this method (with
$<20\%$ accuracy) and charges are
\begin{figure}[!htb]
\begin{center}
\includegraphics*[width=4cm]{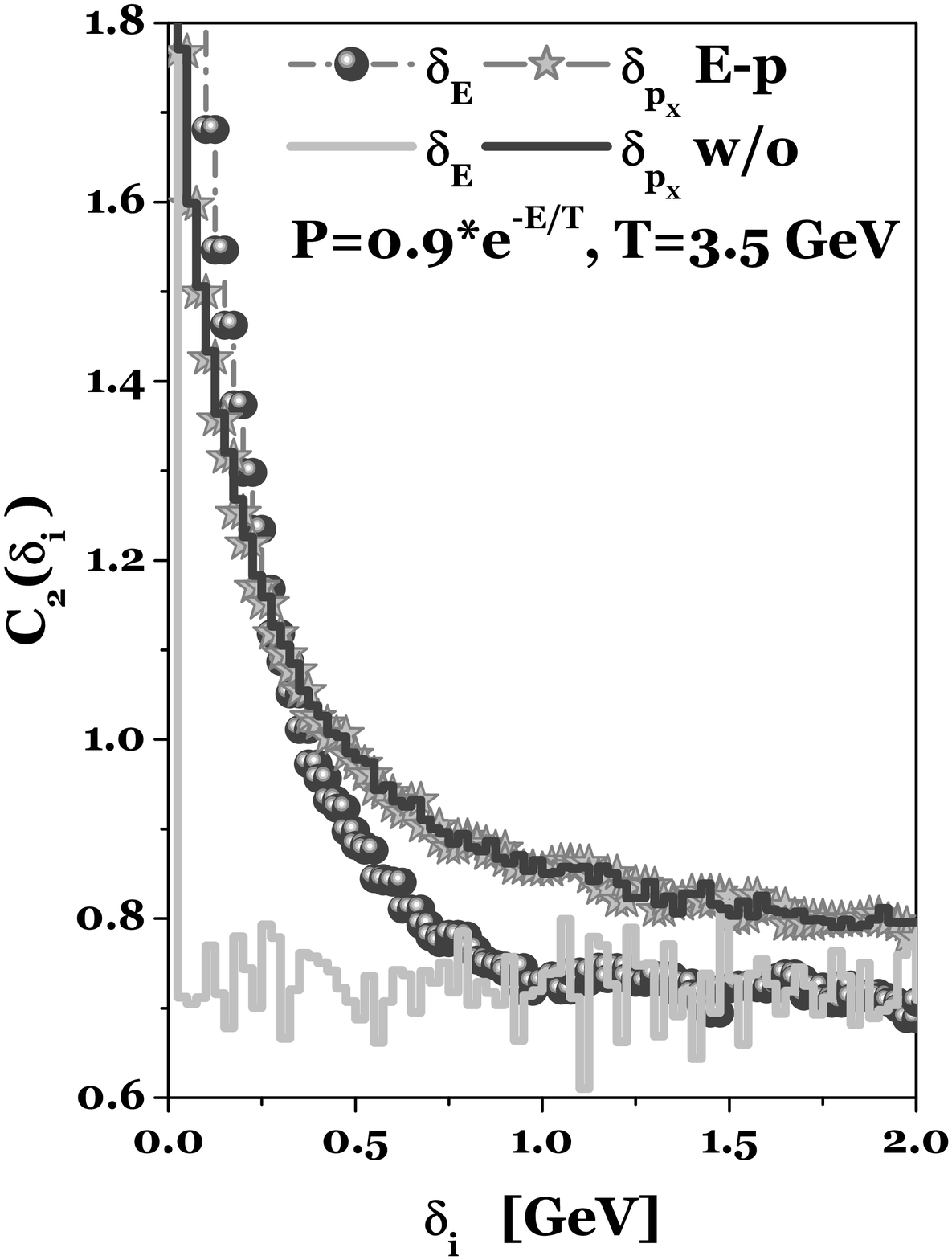}
\includegraphics*[width=4cm]{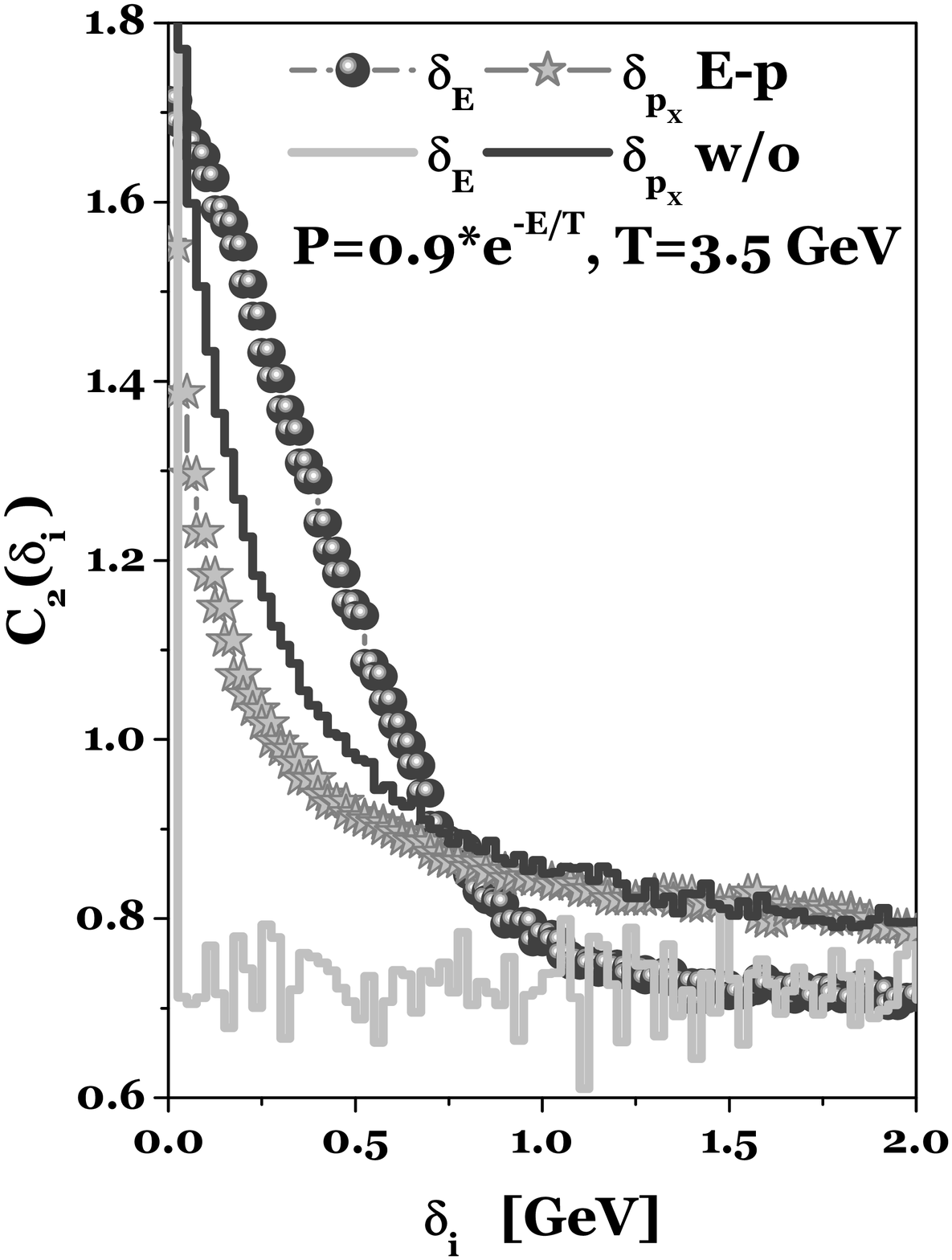}
\includegraphics*[width=4cm]{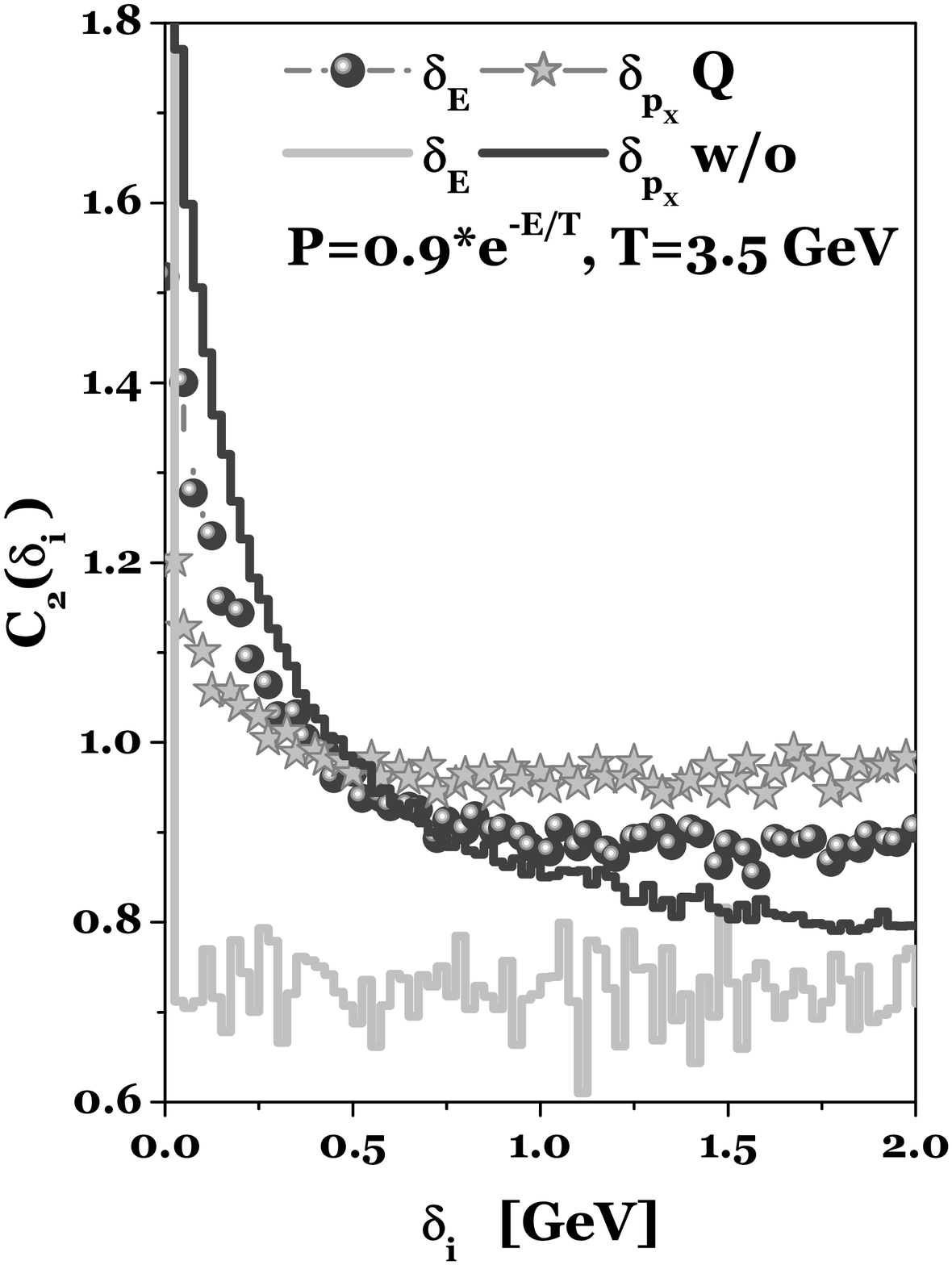}
\includegraphics*[width=4cm]{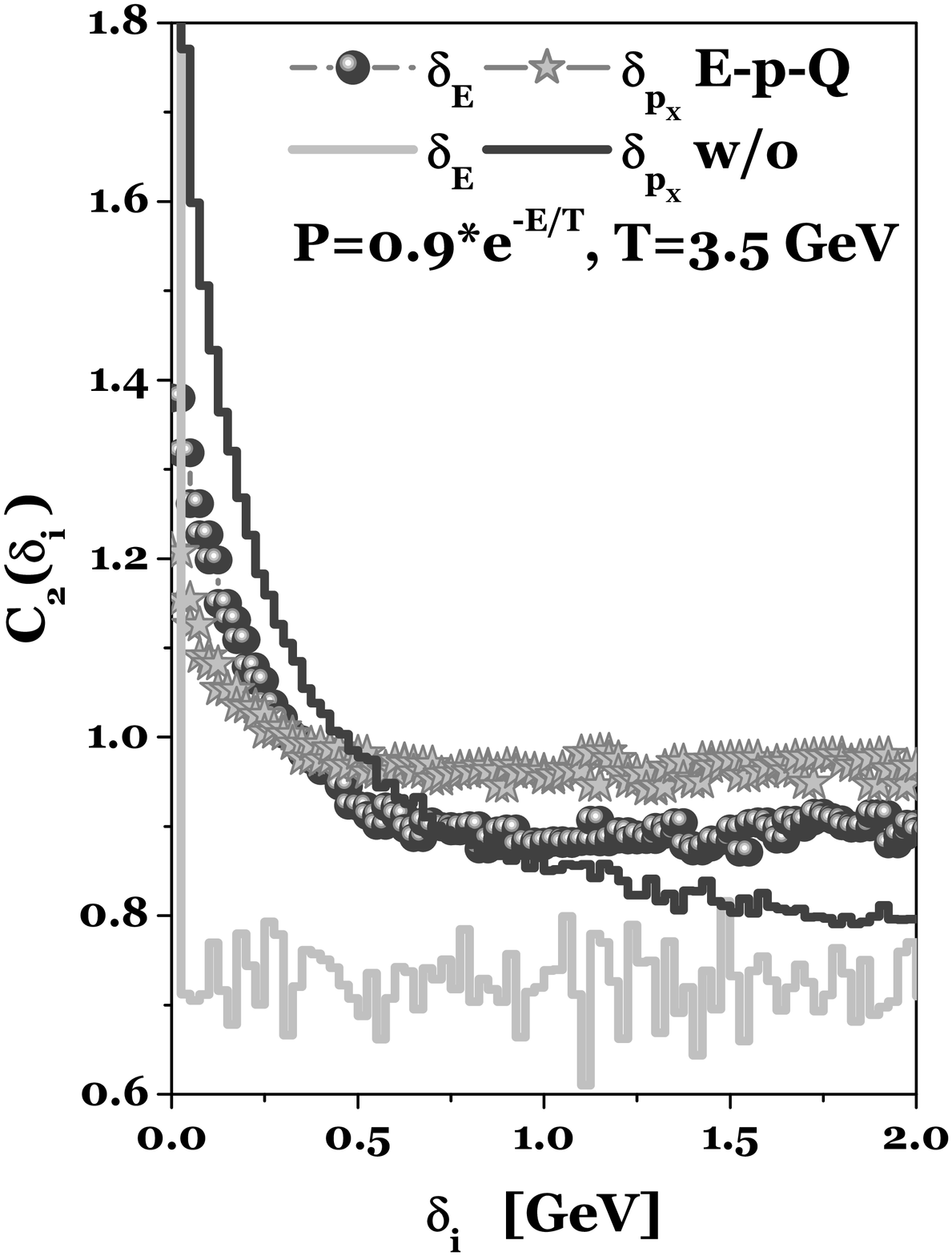}
\end{center}
\vspace{-0.6cm} \caption{Effects of correcting for the imbalances
introduced by our selection procedure in energy-momentum (upper panels
and lower-right panel) and total charge (lower panels). Results without
($w/o$) corrections are the same at all panels. Two selection of the
$(+/-)$ sign of the $p_L$ component are used: it is chosen randomly for
each particle irrespective to EEC it belongs to (all panels except the
upper-left one); all particles in a given EEC possess $p_L$ of the same
sign (upper-left panel). In all panels first bin of gray full curve
contains uncorrected $C_2(\delta_E)$. In all cases $\sigma_0=0$.}
\label{Cons} \vspace{-0.3cm}
\end{figure}
conserved by accepting only events reproducing the initially assumed
charge. Results of all other algorithms presented here were prepared
in the same way as what we have presented, i.e.,  not corrected for
any, therefore they can be compared with each other. We would now
like to discuss changes in correlation functions induced by
correcting for energy-momentum and charge imbalances introduced in
the selection process. At first one must stress that there is no
unique procedure to perform such corrections \cite{EPQ}. In what
follows we shall use a very simple (but fast) procedure: $(i)$ -
shifting (by the same amounts, $\Delta p_x$, $\Delta p_y$ and
$\Delta p_z$) components of momenta, $p_{x,y,z}$ and, after
balancing momenta, appropriately rescaling energies; $(ii)$ by
converting the necessary number of $(+)$ particles to $(-)$ ones (or
vice versa, depending on the actual charge balance in the event)
and, in case of odd number of particles with surplus charge,
attribute the surplus charge to some randomly chosen $(0)$ charged
particle.

\begin{figure}[!htb]
\vspace{0.5cm}
\begin{center}
\includegraphics*[width=6.5cm]{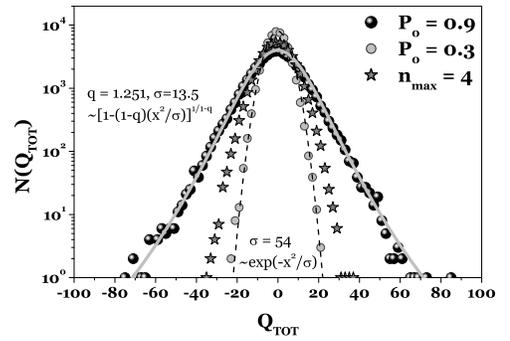}
\end{center}
\vspace{-5mm} \caption{Typical distributions of total charges
obtained from our selection procedure performed for $W=91.2$ GeV,
$T=3.5$ GeV and $\sigma_0 =0.$ for different sizes of EEC as
dictated by parameter $\mathcal{P}_0 = 0.9$ and $\mathcal{P}_0 =0.3$
and for  maximal size of EEC artificially limited to $n_{max} = 4$
(for $\mathcal{P}_0=0.9$). The best fits to most narrow and most
wide distributions are shown. Total initial charge of hadronizing
mass $W$ was assumed $Q=0$.} \label{Q}
\end{figure}

The results are shown in Fig. \ref{Cons} (our hadronizing sources
have zero initial charge). The most sensitive to the correcting
procedure is $C_2(\delta_E)$, correlations of momenta feel
corrections only when all particles in a given EEC are located in
the same hemisphere. Otherwise there is essentially no difference,
The reason for such behavior is that in the other case, the relative
differences of momenta considered here are not changed by the
shifting procedure. In correcting for $\Delta Q\neq 0$ the EECs with
single particle only were chosen first, afterwards EECs with
$N_{part} >1$ were chosen randomly until the correct balance of
charge was achieved. In this case, as one can see in Fig. \ref{Cons}
(lower panels), the effect is quite dramatic and essentially
independent of the way the momenta were chosen or on their
energy-momentum balance. The best measure of this is provided by the
widening of the originally $\delta$-like $C_2(\delta_E)$. Notice
also the clearly visible upward bending of the tail of $C_2$
distributions. It is worth notice at this point that similar shapes
are observed in $C_2$ obtained from $e^+e^-$ annihilation
experiments, in which, as in the case considered here, the original
energy and total charges are well known and fixed.

Effects shown here are so dramatic because, in order to be as near
as possible to the proper BE distributions in the average EEC in
situation of only limited energy $W$ available for hadronization, we
had maximized sizes of EECs by allocating to them many particles,
this was technically achieved by using large value of parameter
$\mathcal{P}_0$, $\mathcal{P}_0=0.9$. It resulted in very broad
total charge distribution (centered on the assumed value
$Q_{TOT}=0$), see Fig. \ref{Q}. The reason for such large
fluctuations is as follows. In our algorithm each EEC contains
particles of the same charge $(+,-,0)$ selected randomly (with equal
probabilities).  With only one particle per EEC, $N_{part}=1$, this
would result in quite narrow $\Delta Q_{EEC}$. However, because in
general $N_{part} > 1$ and fluctuates, the charge imbalance $\Delta
Q_{EEC}$ broadens considerably to what is observed in Fig. \ref{Q}.
It leads then to very large fluctuations and events with large
charge imbalance are quite frequent. The effect is therefore, as
seen in Fig. \ref{Q}, very sensitive to the size of EEC allowed (see
Fig. \ref{fig:cells} and Table \ref{Parameters}), i.e., to the
parameter $\mathcal{P}_0$ responsible for $N_{part}$ and to any
attempts to limit it as, for example, shown in Fig. \ref{multlim}
\cite{FOOTQ}.

To summarize: this problem arises because our procedure destroys
part of the originally formed EECs and forms some new ones what
results in the sensitivity observed. On the other hand we do not see
at the moment any economical way to account for extremely
complicated correlations arising when attempting to keep $Q_{TOT}=0$
all the time. The only apparent cure, to keep only events with
exactly right charge balance, would place our algorithm in the same
category (in what concerns the use of computational time) as those
presented in Sects. \ref{sec:zajc} and \ref{sec:cramer}
\cite{zajc,Cramer} (which, by the way, were not attempting to impose
any conservation laws at all).



\begin{thebibliography}{99}
\bibitem{FOOT4} The only discussion of such changes we are aware of is in
                \cite{UWW1,UWW2} where it was shown that enforcing effect of
                BEC into a simple cascade model results in the appeareances of
                multi-charged vertices, not present in the original scheme.
                For more details see \cite{AIP} and references therein.
\bibitem{FOOT5} The reason is obvious. When only mean multiplicity $\bar{n}$ is
                fixed the most probable distribution according to IT approach is
                geometrical (or Bose-Einstein) one. Such is therefore
                distribution of particles in each cell in rapidity, therefore their
                composition will result in $P(n)$ of NB type \cite{NBD}.
\bibitem{FOOT6a} Notice that, in general, the multiplicity distribution function
                 satisfies the relation: $ P(n+1)/P(n) = g(n)/(n+1)$ where
                 $(n+1)$ is reflection of the fact that with $n$
                 particles given the $(n+1)$th one can be allocated
                 in $n+1$ ways. Different $g(n)$ result in different
                 forms of $P(n)$. In particular for $g(n)= a$ one gets
                 (after normalization) $P(n) = a^n e^{-a}/n!$ whereas
                 $g(n) = p(n+1)$ results (after normalization) in
                 $P(n) = (1-p) p^n$, i.e., in a geometrical distribution,
                 which means that the emission of an extra particle is enhanced
                 by a factor $(n+1)$ and this is Bose-Einstein enhancement
                 and the emission with this is usually called stimulated
                 emission \cite{GN}.
\bibitem{FOOT12} Notice that the problem of how to account for the possibly
                 nonzero spin of produced resonances is highly nontrivial.
                 In principle, one should build EECs with
                 particles possessing either the same helicity (but then
                 they would all have to possess momenta pointing in the
                 same direction) or with the same projection of spins
                 (but then one would have to choose the external axis
                 and choose not energies, as we are doing so far, but
                 rather momenta). In both cases it would result in a
                 completely different approach, which we shall not
                 pursued here.
\bibitem{FOOTQ} It is interesting to note that distributions of $Q_{TOT}$ in Fig.
                \ref{Q} follow Gaussian form for small small $\mathcal{P}_0$, for
                $\mathcal{P}_0=0.9$ which becomes the so called $q$-Gaussian
                distribution (known from Tsallis statistics, see reference in
                \cite{FOOT1}).
\end{thebibliography}
\end{document}